\documentclass[pdflatex,sn-nature]{sn-jnl}
\usepackage{graphicx}%
\usepackage{multirow}%
\usepackage{amsmath,amssymb,amsfonts}%
\usepackage{amsthm}%
\usepackage{mathrsfs}%
\usepackage[title]{appendix}%
\usepackage{xcolor}%
\usepackage{colortbl}
\usepackage{textcomp}%
\usepackage{manyfoot}%
\usepackage{booktabs}%
\usepackage{algorithm}%
\usepackage{algorithmicx}%
\usepackage{algpseudocode}%
\usepackage{listings}%
\usepackage{makecell}%
\usepackage{array}
\usepackage{adjustbox}
\usepackage{pdflscape}
\usepackage{amsmath}
\usepackage[T1]{fontenc}
\usepackage{fancyhdr}
\usepackage[normalem]{ulem}
\fancypagestyle{lscape}{
  \fancyhf{}
  \fancyfoot[C]{\thepage}

  \fancypagestyle{lscape}{
  \fancyhf{}
  \rfoot{\begin{turn}{90}\thepage\end{turn}}

}

}
\newcolumntype{L}[1]{>{\raggedright\arraybackslash}p{#1}}

\theoremstyle{thmstyleone}%
\theoremstyle{thmstyletwo}%

\theoremstyle{thmstylethree}%

\newcommand{\pardir}{\circ \kern-4pt \rightarrow}
\newcommand{\nondir}{\circ\!\!-\!\!\circ}

\newcommand{\pa}{\text{pa}}
\newcommand{\indep}{\perp\!\!\!\perp}

\usepackage{setspace}
\usepackage{geometry}
\begin{document}


\title{Mapping the causal structure of price formation in Texas's transitioning electricity market}

\author*[1]{\fnm{Shiva} \sur{Madadkhani}}\email{shiva.madadkhani@tum.de}

\author[3]{\fnm{Nils} \sur{Sturma}}\email{nils.sturma@epfl.ch}

\author[2]{\fnm{Mathias} \sur{Drton}}\email{mathias.drton@tum.de}

\author[1]{\fnm{Svetlana} \sur{Ikonnikova}}\email{svetlana.ikonnikova@tum.de}

\affil[1]{\orgdiv{School of Management}, \orgname{Technical University of Munich},  \country{Germany}}

\affil[2]{\orgdiv{School of Computation, Information and Technology}, \orgname{Technical University of Munich and Munich Center for Machine Learning},  \country{Germany}}

\affil[3]{\orgdiv{Insitute of Mathematics}, \orgname{École Polytechnique Fédérale de Lausanne},  \country{Switzerland}}


\abstract{
Renewable deployment and rising demand from electrification and large digital loads are transforming electricity markets. However, how these developments reshape electricity price dynamics remains poorly understood, leaving system planners, capacity investors, and market participants reliant on assumptions from a thermal-dominated era that may no longer hold. We use causal discovery to study the evolution of wholesale electricity prices in Texas, which is undergoing rapid transformation. Our findings overturn the view of Texas as a gas-price-driven market, demonstrating that wind generation has become the dominant causal driver of day-ahead prices, with effects more than three times greater than those of natural gas. Yet wind's price-suppressing effect is weakening during peak periods, and wind growth redistributes congestion costs to distant load centres. Furthermore, rising load in South and West Texas alters system prices and regional differentials. Uncovering the evolving spatiotemporal nature of causal drivers, our analysis reveals that the pace, geographic siting, and relative scale of new generation and large loads will be decisive for future electricity price risks, infrastructure needs, and investments.}




\maketitle
\doublespacing
\renewcommand{\thefootnote}{\fnsymbol{footnote}}
\footnotetext[1]{
Shiva Madadkhani and Nils Sturma acknowledge support by the Munich Data Science Institute (MDSI) at Technical University of Munich (TUM) via the Linde/MDSI PhD Fellowship program in 2021-2024. This project has received funding from the European Research Council (ERC) under the European Union’s Horizon 2020 research and innovation programme (grant agreement No 883818). We thank participants at the 46\textsuperscript{th} IAEE International Conference in Paris for their helpful comments on an earlier version of this work, which appears in the \href{https://iaee2025paris.org/download/contribution/abstract/828/828_abstract_20250409_162652.pdf}{conference proceedings}. We also thank Moritz Ebert for helpful preliminary work conducted in his master's thesis, and are grateful to Gunther Glenk, G\"urcan G\"ulen and Sebastian Schwenen for helpful suggestions.
}

\renewcommand{\thefootnote}{\arabic{footnote}}

\clearpage
\doublespacing
\section*{Main}\label{sec1}

The evolution of electricity price dynamics shapes investment decisions, infrastructure planning, and market participant behaviour, with consequences for reliability and affordability of future electricity supply. Yet our conventional understanding of electricity price formation, inherited from an era of centralised, thermal-dominated power systems, is called into question by rapid structural changes on both sides of the market: large-scale renewable integration, shifting demand patterns \cite{bistline2021deep, shaffer2022changing}, and the emergence of new large loads such as data centres \cite{nerc_large_loads_2025, norris_load_2025} are transforming global electricity systems. Through co-located generation, behind-the-meter storage, and long-term contracts with renewable developers, large electricity consumers increasingly influence supply-side infrastructure decisions, while their own location choices are in turn shaped by prices and generation availability \cite{heiss2026powering}. Without a clear understanding of what these ongoing developments imply for price dynamics, decision-makers may misinterpret price signals, make misguided investments, misallocate infrastructure, and ultimately expose consumers to unnecessarily higher costs.

Undergoing radical growth on both supply and demand sides, Texas presents a salient example of fast-paced power sector transformation in the United States. Population growth, industrial expansion, electrification of the oil and gas industry, and a recent surge in cryptocurrency operations and data centres \cite{peter_r_hartley_ercot_2024, tyler_hodge_data_2024} have raised electricity consumption in Texas by 114 TWh (33\%) over the decade to 2024 \cite{eia_steo_2026}. Meanwhile, natural resource availability, falling technology costs, and a supportive regulatory landscape have spurred substantial growth in wind and solar deployment, making Texas the largest producer of renewable electricity in the United States \cite{peter_r_hartley_ercot_2024, eia_steo_2026}. As electricity supply and demand have expanded, their spatial distributions and temporal profiles have become increasingly mismatched \cite{peter_r_hartley_ercot_2024}, leading to complex, regionally differentiated price dynamics across the state (Fig. \ref{figure1}). These Texas developments and their price consequences offer a preview for other regions undergoing rapid transformation and facing shifting demand patterns from artificial intelligence (AI) and electrification. 

Although the literature on electricity price dynamics is extensive, ranging from studies of the merit-order effect \cite{woo2011impact, woo2011wind, ketterer2014impact, clo2015merit, rintamaki2017does, ajanaku2024comparing} to deep learning approaches for electricity price forecasting \cite{jkedrzejewski2022electricity}, it almost entirely relies on associational patterns in historical data. However, rapid changes in both supply- and demand-side drivers undermine the applicability and predictive power of such models. This makes identifying causal relationships essential, since they distinguish causal drivers from spurious patterns or mere co-movements that associational (i.e., non-causal) approaches cannot separate \cite{peters2016causal, cacciarelli2025we}, and are therefore more robust to structural change. Revealing the directionality of the relationships becomes even more important as supply and demand increasingly become entangled. Causal analyses that could shed light on evolving electricity market linkages, however, remain comparatively limited. Past studies have applied causal analysis to estimate demand elasticities \cite{tiedemann2024identifying}, or to examine linkages between fuel and electricity prices \cite{mjelde2009market, ferkingstad2011causal, nakajima2013testing}, across interconnected electricity markets \cite{park2006price, ferkingstad2011causal, castagneto2014dynamic}, and from individual drivers such as renewable generation to prices \cite{cacciarelli2025we}. Yet existing studies have not accounted for the latent confounding and autocorrelation ubiquitous in power system data when discovering causal structure, which can result in misidentification of causal relationships \cite{glymour2019review, pearl2009causality, spirtes2000causation, peters2017elements, shojaie2022granger, castagneto2014dynamic, runge2023causal, runge2019detecting}. What is missing is a causal analysis that addresses these methodological issues and systematically examines how interactions among major drivers evolve, reshaping electricity price patterns over time. 

Here, we employ recent advances in causal discovery to examine causal relationships among a rich set of electricity price determinants, including renewable generation, natural gas dynamics at three major gas hubs, and regional demand patterns across Texas over 2019--2024. We demonstrate that wind generation has overtaken natural gas prices as the dominant causal driver of day-ahead electricity prices, with effects more than three times larger, challenging the view of Texas as a gas-price-driven power market. Wind's price-suppressing effect, however, is weakening during the grid's stressed periods, and wind growth leads to redistribution of congestion costs to distant load centres in seasonally shifting patterns. Natural gas prices remain causally relevant for electricity prices, though the influential hub has switched from regional Waha to the national Henry Hub. On the demand side, South and West Texas loads gain influence as electrification and data centre growth reshape regional consumption, further contributing to how demand-side drivers shift over time. Estimating the strength of causal linkages, we reveal the spatiotemporal nature of market signal propagation. Therefore, when, where, and at what scale new generation and large loads are sited will be crucial for future price risks, infrastructure needs, and the allocation of investment across the grid.

\subsection*{Mapping price dynamics in ERCOT}\label{sec2}

We focus on day-ahead market (DAM) prices at the four main trading hubs (North, South, West, and Houston) in the Electric Reliability Council of Texas (ERCOT), which serves 90\% of the state's load \cite{du2023renewable}. We decompose DAM prices into two components: the system lambda ($\lambda$), which reflects the system-wide energy component of prices and is uniform across hubs, and hub-specific price differentials ($\Delta_h$) that aggregate node-level congestion effects within each hub $h$ (Fig. \ref{figure2}; see \nameref{sec:methods} for a brief introduction to ERCOT's day-ahead pricing). This decomposition allows us to separate the causal drivers of system-wide prices from regional price differentials. We estimate separate causal graphs for ERCOT-defined peak (06:00–22:00) and off-peak hours, and for warmer (May--October) and cooler (November--April) months, resulting in four distinct market regimes that reflect differences in demand patterns, renewable output, and natural gas market conditions. To track how causal relationships evolve over time, we implement a rolling two-year window over the 2019--2024 period. This yields 20 regime‑specific causal graphs (four regimes × five windows), allowing us to identify which relationships are stable, which vary with system and market conditions, and which undergo structural change.

\begin{figure}[htbp]
    \centering
    \includegraphics[width=0.8\textwidth]{fig1.pdf}
    \caption{Geographic distribution of generation capacity, natural gas infrastructure, and demand centres across Texas. Power plant locations are shown by fuel type (wind, solar, natural gas), with circle size proportional to installed capacity (MW). Blue shading indicates counties with wind production; grey shading denotes major oil-producing counties (total crude oil production $>$ \$15 million barrels, January–December 2024). Key demand centres, the Waha and Houston Ship Channel natural gas hubs are labelled. Map shapefile from Texas Open Data Portal \cite{texas_counties_map}; Power plant data from EIA \cite{eia860}; Oil-producing counties from the Railroad Commission of Texas \cite{rrc_pdq}.}
    \label{figure1}
\end{figure}

Our variable selection is guided by the literature on electricity price determinants and our own previous work \cite{woo2011impact, zarnikau2019market, madadkhani2024toward} (see \nameref{sec:methods} for the full variable list). The set of variables influencing electricity prices is prohibitively large, making the omission of relevant common causes unavoidable. This violates a core assumption of standard causal discovery methods, which require that all common causes are observed, leading them to mistake confounded associations for direct causal effects \cite{spirtes2000causation}. This problem is compounded by the strong autocorrelation inherent in power system variables, under which standard causal discovery methods overinflate or falsely detect causal effects \cite{runge2019detecting}.  To address these challenges, we employ the Latent Peter-Clark Momentary Conditional Independence (LPCMCI) algorithm \cite{gerhardus2020high-recall}, a recent advance in time series causal discovery designed to handle latent confounders and autocorrelation (see \nameref{sec:methods}). We include time lags of up to seven days to capture both contemporaneous (within-day) and delayed (between-day) causal effects, reflecting the well-documented weekly seasonality of electricity demand and prices \cite{weron_electricity_2014}.

The LPCMCI algorithm produces a matrix of statistical test results that identifies which observed associations can be interpreted as causal and determines their direction. We visualize these results as causal graphs, in which nodes represent variables and edges represent directed causal relationships (Fig. \ref{figure2}). Edge colours reflect the strength of conditional associations (i.e., partial correlations). A directed edge from $X \to Y$ indicates that intervening on $X$ would cause a change in $Y$. A bidirected edge ($X\leftrightarrow Y$) indicates that the association between two variables is best explained by an unobserved common cause rather than a direct causal link. Edges with circles on one or both ends signal that an association is present but the available data do not allow us to determine the causal direction. The absence of an edge indicates evidence against a causal relationship (see \nameref{sec:methods}). 

As Fig. \ref{figure2} illustrates for one window, each inferred causal graphs reveals a dense network of relationships between electricity prices and their fundamental determinants. To facilitate interpretation, we organize the results into four blocks: supply-side drivers, demand-side drivers, natural gas price dynamics, and day-ahead price components. We then trace how causal effects propagate to electricity prices and natural gas (NG) generation volumes, which we include throughout as causal effects on prices may operate through changes in NG-fired dispatch.

\begin{figure}[htbp]
    \centering
    \includegraphics[width=0.65\textwidth]{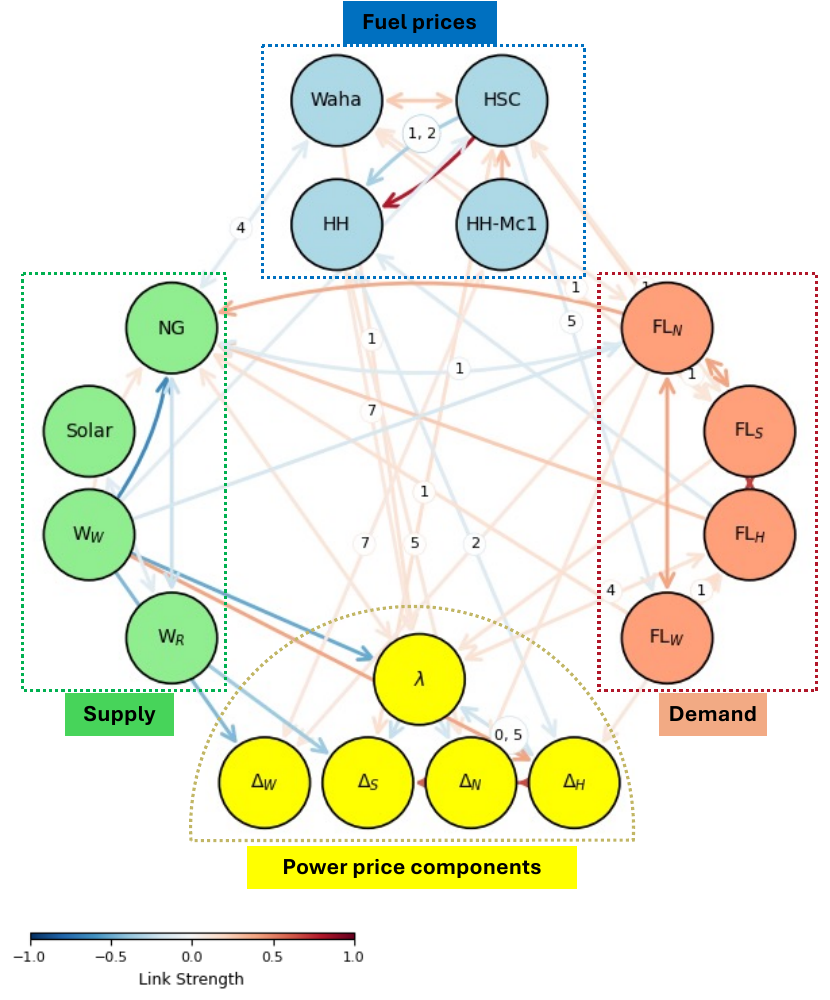}
    \caption{Causal graph for peak hours during cooler months, 2019--2020. Nodes represent variables and edges represent causal relationships inferred by the LPCMCI algorithm (see Methods for a detailed description of edge types). Control variables (see Methods, Table 1) are excluded for clarity. To aid interpretation, we organise variables into four blocks—supply-side drivers, demand-side drivers, natural gas prices, and day-ahead price components—and examine each in turn in the Results. Edges without a lag label denote contemporaneous relationships (lag 0); numbered edges indicate the lag in days.}
    \label{figure2}
\end{figure}

\subsection*{Electricity prices and renewable generation} 

\begin{landscape}

\begin{figure}[htbp]
\rfoot{\begin{turn}{90}\thepage\end{turn}}
\centering
\vspace*{-0.5cm}

\newcommand{\colw}{0.13\linewidth}
\newcommand{\headersize}{\scriptsize}
\newcommand{\rowlabelsize}{\scriptsize}

\setlength{\tabcolsep}{1pt}
\renewcommand{\arraystretch}{0.3}

\begin{tabular}{c ccccc}
 & \headersize 2019--2020 & \headersize 2020--2021 & \headersize 2021--2022 & \headersize 2022--2023 & \headersize 2023--2024 \\[2pt]

\raisebox{.9cm}{\rotatebox{90}{\rowlabelsize cool peak}} &
\includegraphics[width=\colw]{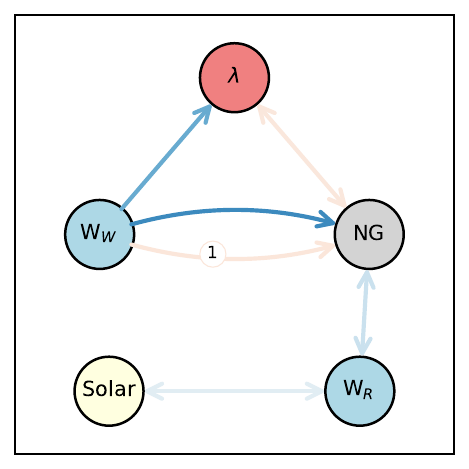} &
\includegraphics[width=\colw]{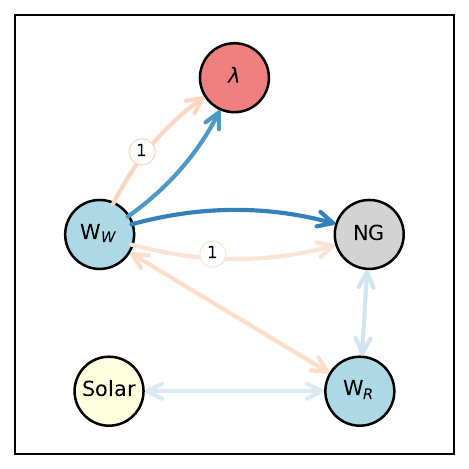} &
\includegraphics[width=\colw]{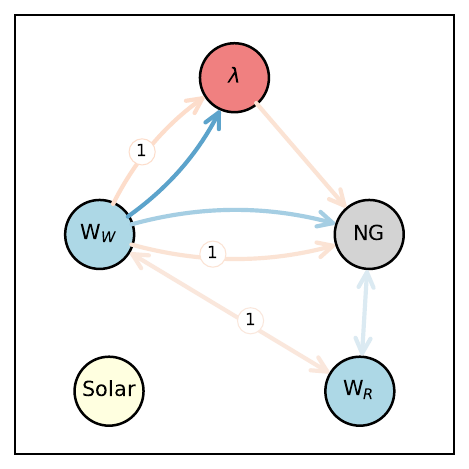} &
\includegraphics[width=\colw]{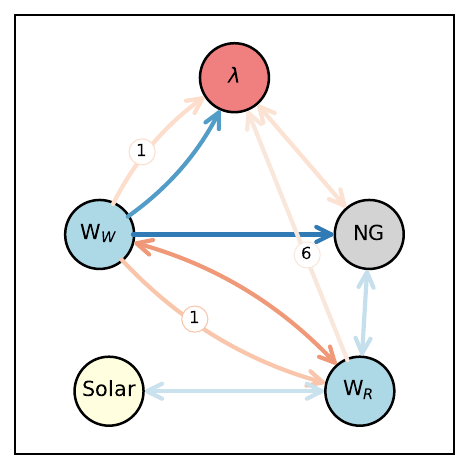} &
\includegraphics[width=\colw]{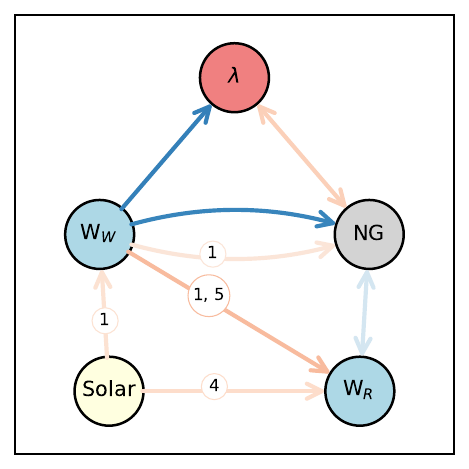} \\

\raisebox{.8cm}{\rotatebox{90}{\rowlabelsize cool off-peak}} &
\includegraphics[width=\colw]{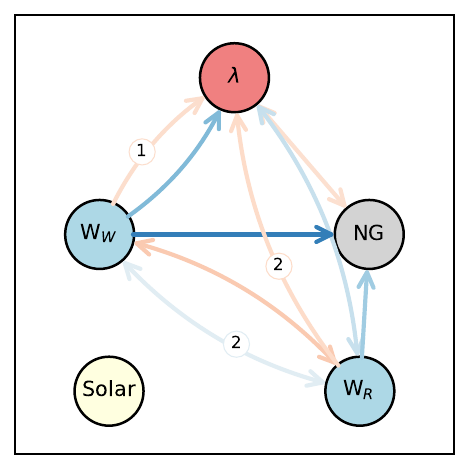} &
\includegraphics[width=\colw]{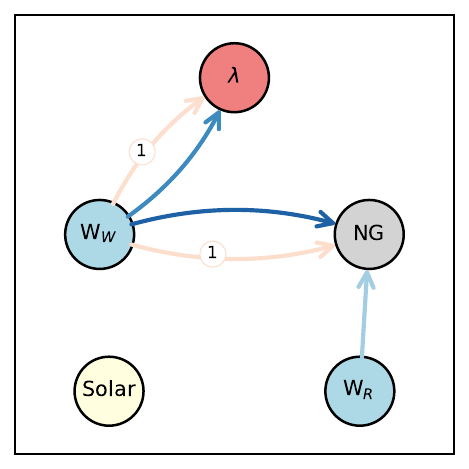} &
\includegraphics[width=\colw]{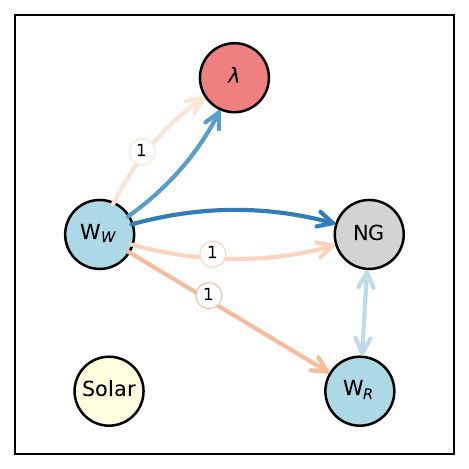} &
\includegraphics[width=\colw]{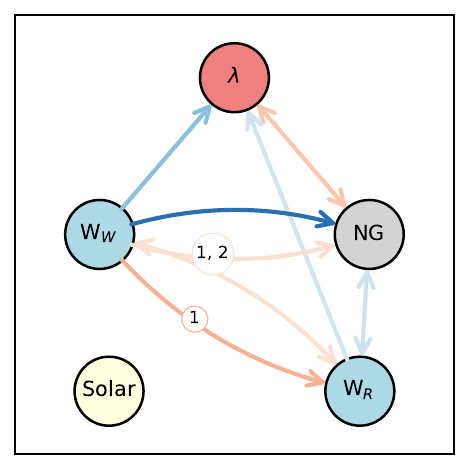} &
\includegraphics[width=\colw]{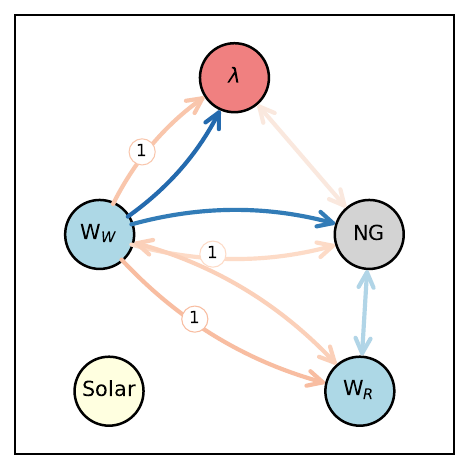} \\

\raisebox{.9cm}{\rotatebox{90}{\rowlabelsize warm peak}} &
\includegraphics[width=\colw]{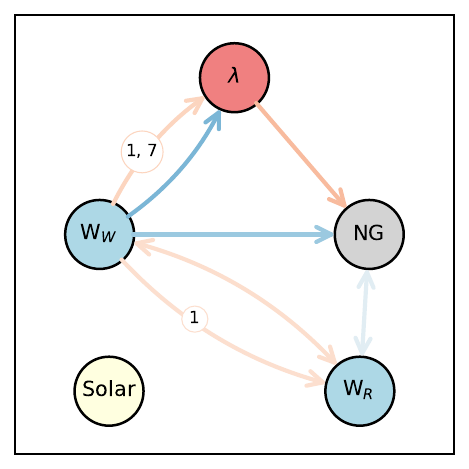} &
\includegraphics[width=\colw]{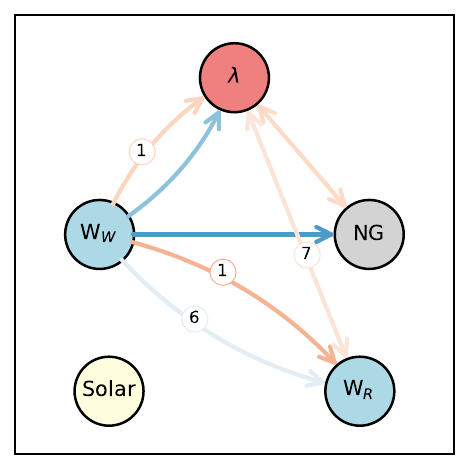} &
\includegraphics[width=\colw]{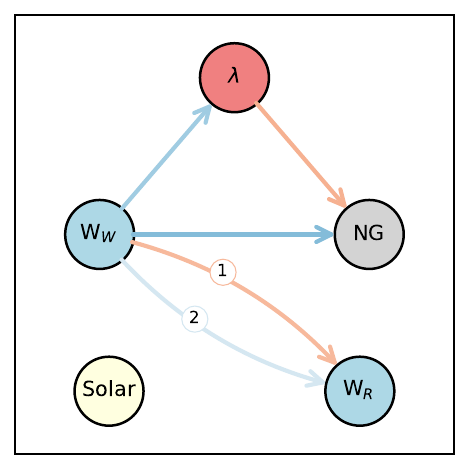} &
\includegraphics[width=\colw]{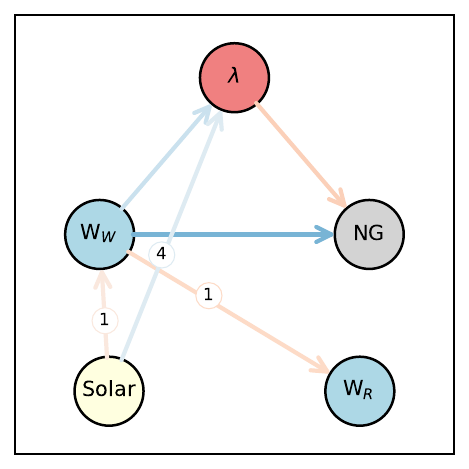} &
\includegraphics[width=\colw]{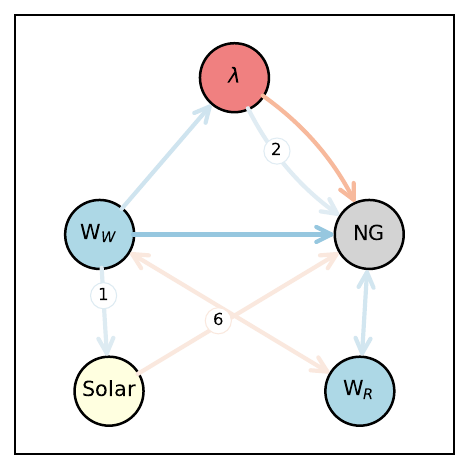} \\

\raisebox{.7cm}{\rotatebox{90}{\rowlabelsize warm off-peak}} &
\includegraphics[width=\colw]{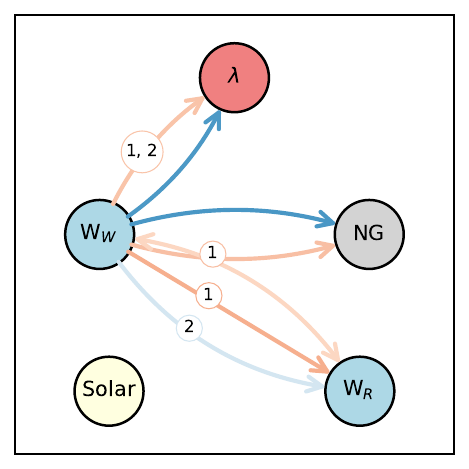} &
\includegraphics[width=\colw]{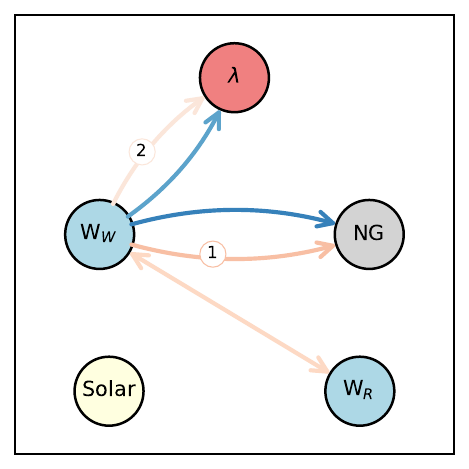} &
\includegraphics[width=\colw]{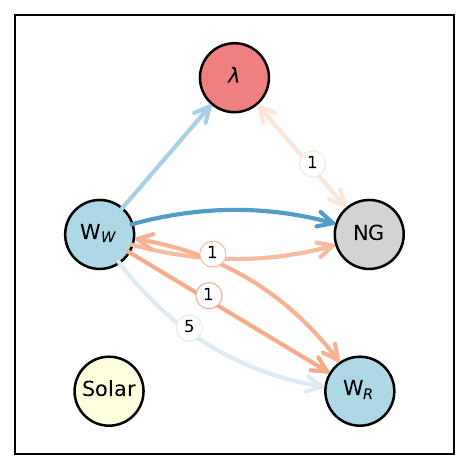} &
\includegraphics[width=\colw]{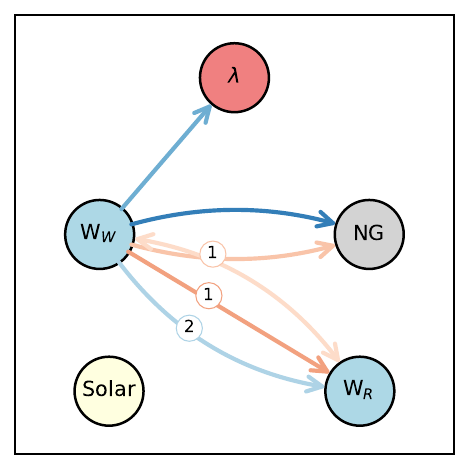} &
\includegraphics[width=\colw]{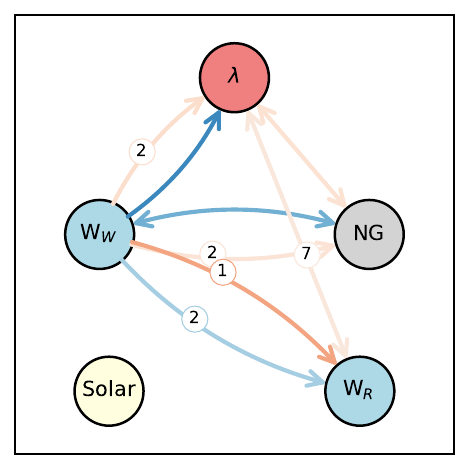} \\ [6pt]
 & \multicolumn{3}{l}{\includegraphics[width=0.21\linewidth]{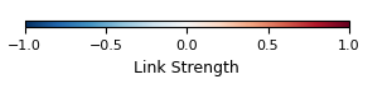}} \\

\end{tabular}

\caption{
    Causal effects of renewable generation on system lambda and natural gas generation. Each panel shows the causal graph for a distinct market regime (peak/off-peak, warm/cool) over successive two-year windows from 2019 to 2024. W$_W$ denotes forecast wind generation in West Texas, W$_R$ forecast wind in the rest of Texas, and Solar forecast solar generation. Edge colour indicates the sign and magnitude of the causal effect (see colour scale). Edges without a lag label denote contemporaneous relationships (lag 0); numbered edges indicate the lag in days. Autoregressive edges (self-loops) \textbf{are omitted for clarity.}}
    \label{figure3}
\end{figure}
\end{landscape}

For the system lambda component of prices, West Texas wind forecast (Wind$_{west}$) emerges as the strongest causal driver across all periods analysed, exerting a consistently negative effect on both lambda and NG generation (Fig. \ref{figure3}). This likely reflects the concentration of wind capacity in West Texas and its displacement of higher-cost fossil generation. The causal effect of Wind$_{west}$ on NG generation is stronger in cooler than warmer periods in both peak and off-peak hours, indicating greater displacement of gas units in winter. The price-dampening effect on system lambda in peak hours is likewise stronger in cooler months, with the seasonal gap widening over time, as the cool-period effect strengthens modestly (by 35\%) while the warm-period effect roughly halves (Supplementary Fig. \ref{fig:figureS1}). Aggregate wind curtailment rates have not risen in warm months over the study period (Supplementary Fig. \ref{fig:figureS2}), arguing against curtailment as the explanation. Instead, wind generation and transmission capacity additions may not be keeping pace with the scale and geographic pattern of demand growth during warm peak hours, when the system is most stressed.

Beyond same-day effects, Wind$_{west}$ shows a small positive causal effect at a one-day lag on NG generation, and occasionally on lambda, particularly outside warm peak hours. One possible explanation is unit commitment dynamics: high wind output in West Texas can trigger shutdowns of combined-cycle gas plants \cite{FellRoh2019}, which dominate Texas's gas fleet. Decommitted units cannot restart quickly, as 92\% of Texas's combined-cycle plants require 1–12 hours or longer to reach full load from shutdown \cite{eia860}, and may therefore not be recommitted until the following day.

By contrast, forecasted wind outside West Texas (Wind$_{non-west}$) and forecasted solar generation are not direct causal drivers of system lambda over the study period. Wind$_{non-west}$ does, however, exhibit causal links to NG generation during cooler periods. Specifically, it reduces NG generation during cool off-peak hours in 2019--2021, reflecting displacement of inframarginal NG units, leaving system lambda unaffected. In other cool periods, both Wind$_{non-west}$ and NG generation appear to be influenced by unobserved common drivers (for example, weather fronts), rather than causing each other directly. These patterns are largely absent in warm periods, appearing in only 2 of 10 tested windows (Fig. \ref{figure3}k-t), consistent with lower seasonal wind output. Solar generation shows no meaningful causal relationships with either system lambda or NG generation (Fig. \ref{figure3}a–e; k-o). Although solar capacity has expanded rapidly in Texas, it still accounts for only 8\% of generation \cite{eia_elec_browser}, and peak-hour windows include early morning and evening periods with limited solar output.

\begin{landscape}
\begin{figure}[htbp]
\centering
\vspace*{-0.5cm}

\newcommand{\colw}{0.13\linewidth}
\newcommand{\headersize}{\scriptsize}
\newcommand{\rowlabelsize}{\scriptsize}

\setlength{\tabcolsep}{1pt}
\renewcommand{\arraystretch}{0.3}

\begin{tabular}{c ccccc}
 & \headersize 2019--2020 & \headersize 2020--2021 & \headersize 2021--2022 & \headersize 2022--2023 & \headersize 2023--2024 \\[2pt]

\raisebox{.9cm}{\rotatebox{90}{\rowlabelsize cool peak}} &
\includegraphics[width=\colw]{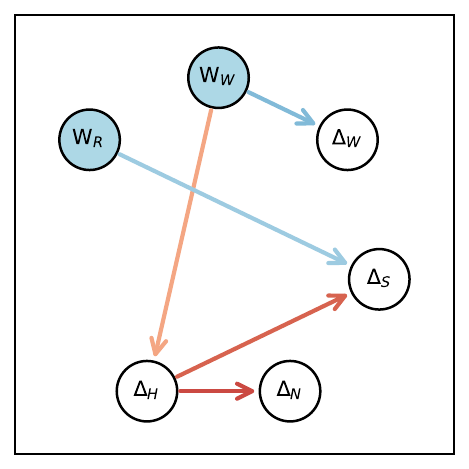} &
\includegraphics[width=\colw]{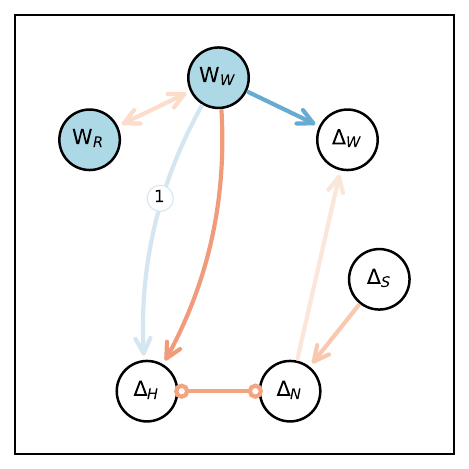} &
\includegraphics[width=\colw]{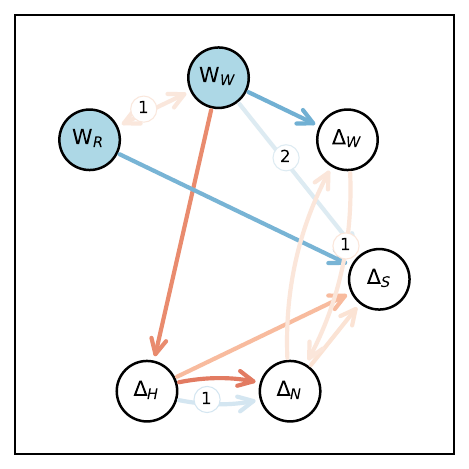} &
\includegraphics[width=\colw]{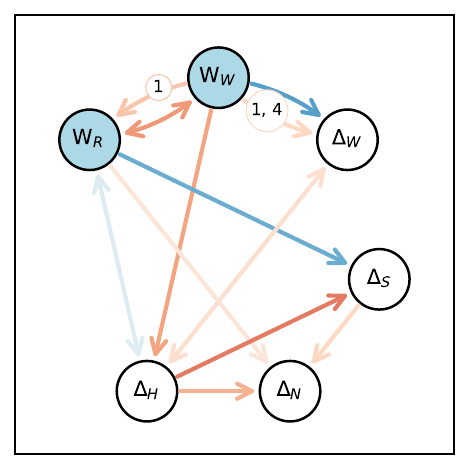} &
\includegraphics[width=\colw]{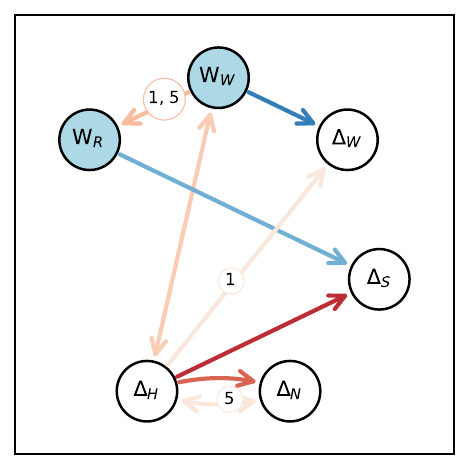} \\

\raisebox{.8cm}{\rotatebox{90}{\rowlabelsize cool off-peak}} &
\includegraphics[width=\colw]{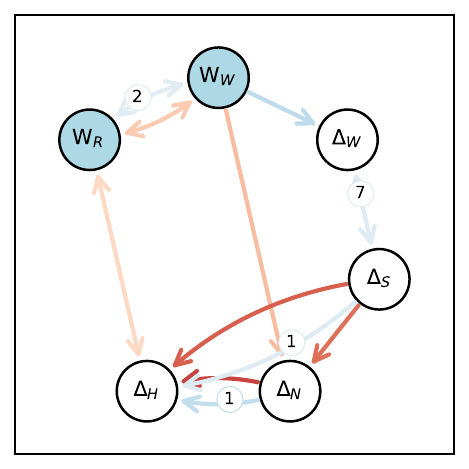} &
\includegraphics[width=\colw]{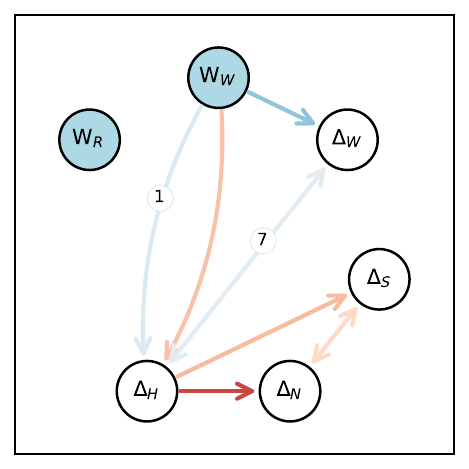} &
\includegraphics[width=\colw]{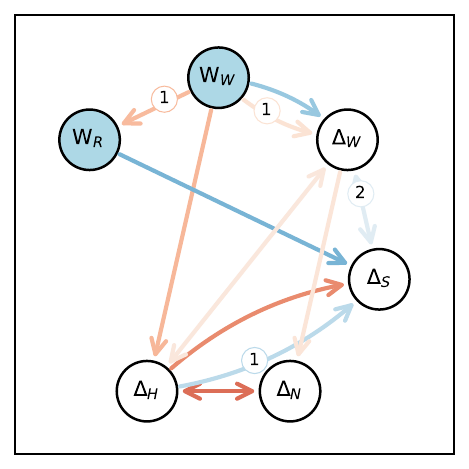} &
\includegraphics[width=\colw]{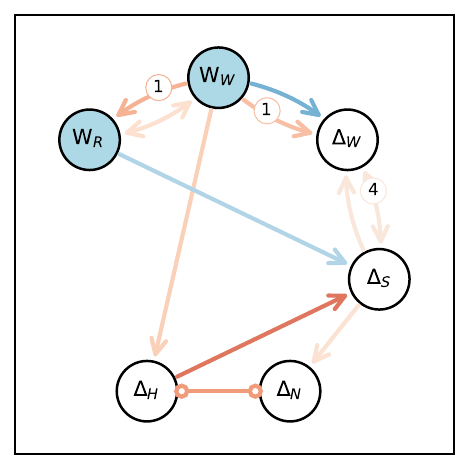} &
\includegraphics[width=\colw]{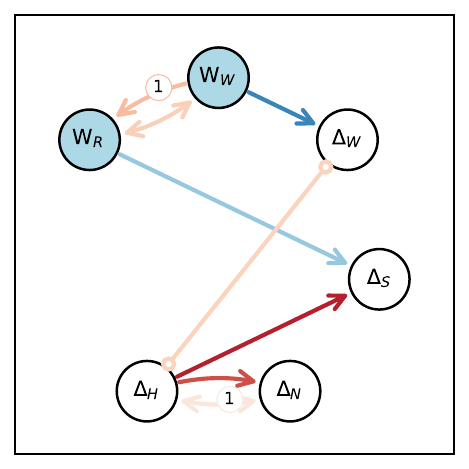} \\

\raisebox{.9cm}{\rotatebox{90}{\rowlabelsize warm peak}} &
\includegraphics[width=\colw]{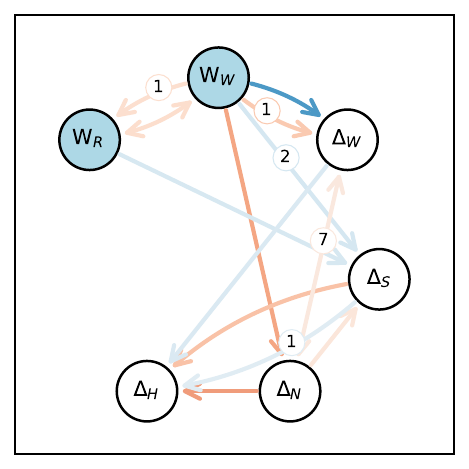} &
\includegraphics[width=\colw]{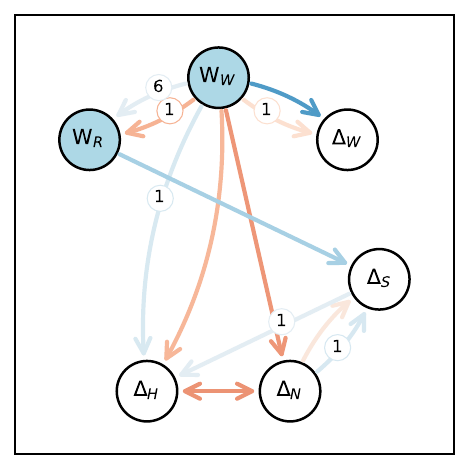} &
\includegraphics[width=\colw]{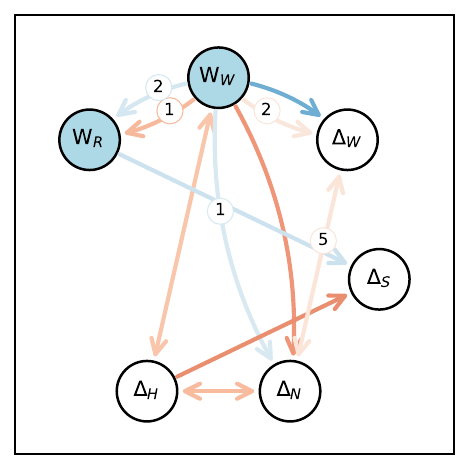} &
\includegraphics[width=\colw]{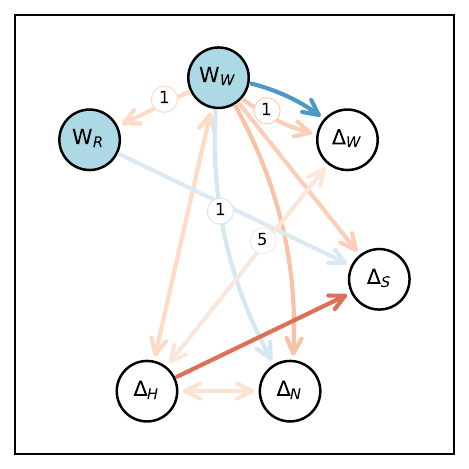} &
\includegraphics[width=\colw]{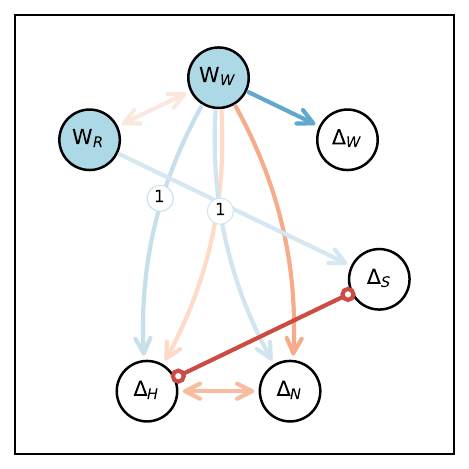} \\

\raisebox{.7cm}{\rotatebox{90}{\rowlabelsize warm off-peak}} &
\includegraphics[width=\colw]{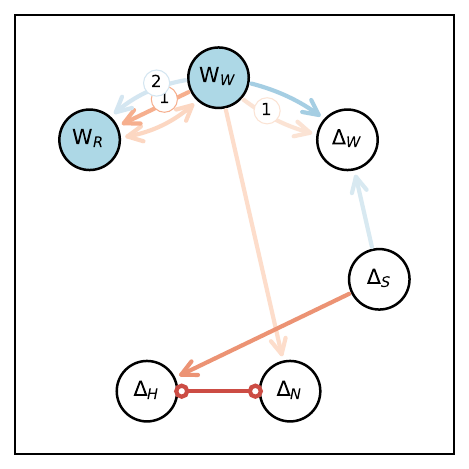} &
\includegraphics[width=\colw]{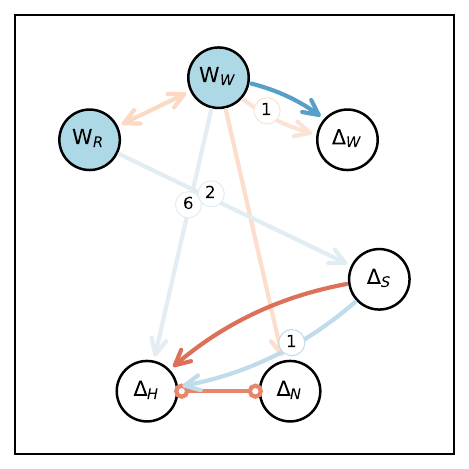} &
\includegraphics[width=\colw]{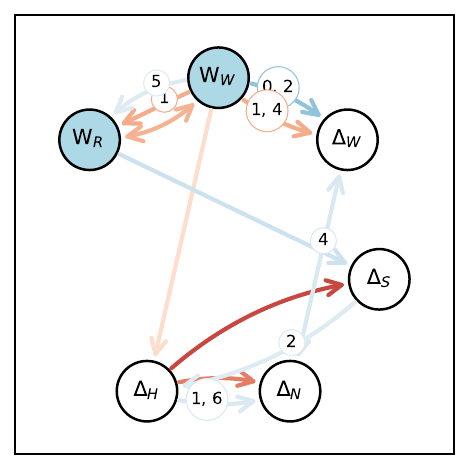} &
\includegraphics[width=\colw]{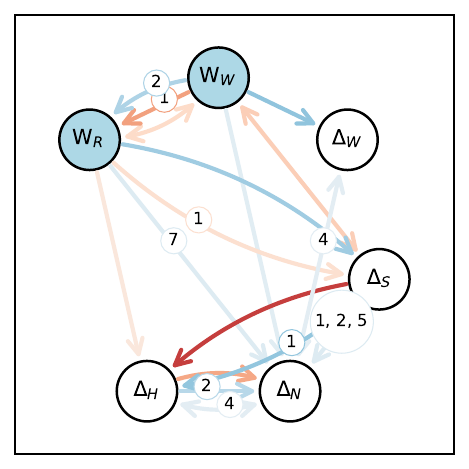} &
\includegraphics[width=\colw]{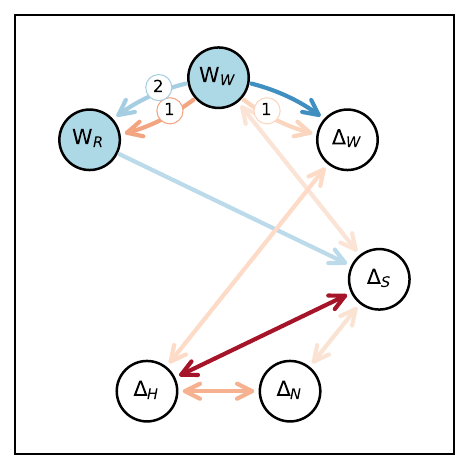} \\[6pt]
 & \multicolumn{3}{l}{\includegraphics[width=0.21\linewidth]{figs/colorbar.pdf}} \\

\end{tabular}

    \caption{Causal relationships between forecast renewables and regional price differentials. $\Delta_N$, $\Delta_S$, $\Delta_W$, and $\Delta_H$ denote the congestion component of day-ahead prices in the North, South, West and Houston price hubs, respectively. W$_W$ denotes forecast wind generation in West Texas, W$_R$ forecast wind in the rest of Texas, and Solar forecast solar generation. Panels are arranged by market regime (peak/off-peak, warm/cool) over successive two-year windows from 2019 to 2024. Edge colour indicates the sign and magnitude of the causal effect (see colour scale). Edges without a lag label denote contemporaneous relationships (lag 0); where a lag is present, the number indicates the delay in days. Autoregressive edges (self-loops) \textbf{are omitted for clarity.}}
    \label{figure7}
\end{figure}
\end{landscape}

For hub price differentials, wind exerts spatially heterogeneous, and sometimes opposing, effects across hubs. It reduces price differentials locally, but its effects on differentials at other hubs vary in direction and magnitude across seasons and times of day (Fig. \ref{figure7}). West Texas wind lowers $\Delta_{\text{West}}$ across all periods, indicating reduced reliance on higher-cost NG generation in the region. However, the effects propagate differently across the system. During cool-season peak hours,  Wind$_{\text{west}}$ increases $\Delta_{\text{Houston}}$, in line with \cite{FellRoh2019}, who show that reduced NG generation, particularly in North Texas, requires Houston to rely more on costlier local units. In contrast, during warm-season peak hours, when NG generation in North continues to operate due to high system-wide demand, Wind$_{\text{west}}$ instead increases $\Delta_{\text{North}}$, pointing to transmission capacity constraints limiting power flows from West to North Texas. During cool off-peak hours, Wind$_{west}$ increases price differentials in both North and Houston by driving the hub that propagates effects to the other in a given period (except the final period). Meanwhile, non-west wind consistently lowers $\Delta_{\text{South}}$ during peak hours (9 of 10 periods) and across all off-peak periods post-2021, indicating that local wind generation reduces South Texas's reliance on more expensive NG generation (Fig. \ref{figure7}).

\subsection*{Regional demand and congestion redistribution}

We next examine the causal effects of regional electricity loads on system lambda. To capture geographic heterogeneity, we distinguish load by region: North, West, South, and Houston. As with renewable generation, we use forecasted rather than realised load, reflecting the information available to market participants at the time of DAM bidding. North Texas forecast load (FL$_{North}$) emerges as the most consistent demand-side driver, with causal effects on system lambda and/or NG generation in 15 of 20 periods (Fig. \ref{figure4}). During warm peak hours, when cooling demand is highest across ERCOT, loads in both North and Houston, the two most populous regions, drive NG generation and/or system lambda. During warm off-peak periods, North and Houston loads continue to show causal links to NG generation, but their influence on system lambda largely disappears. This result is consistent with the system being capacity unconstrained during warm off-peak hours, such that additional natural gas generation meets incremental demand without shifting the marginal price setter.

During cooler periods, South Texas load (FL$_{South}$) plays a distinct and strengthening role in price formation (Fig. \ref{figure4}). It causally drives system lambda and NG generation in all 10 cool-period windows across both peak and off-peak hours, affecting either lambda or NG generation during 2019--2021, and both simultaneously from 2021 onward. This pattern reflects transmission dynamics and the spatial distribution of generation: as FL$_{South}$ increases, less Wind$_{non-west}$ is available for export to North Texas and Houston, either through local absorption or increased congestion in South Texas \cite{potomaceconomics2024}. These regions then rely more heavily on natural gas generation, raising both NG generation and the system marginal cost. This aligns with the earlier finding that Wind$_{non-west}$ shows stronger causal links to NG generation during cooler periods.

Forecasted West Texas load (FL$_{west}$) exhibits a non-monotonic temporal pattern that sets it apart from other regions: it causally affects system lambda during cool periods in 2019--2020, becomes entirely disconnected in 2020--2021, and re-emerges from 2021 onward, most prominently during cold off-peak hours (Fig. \ref{figure4}). This likely reflects interactions between abundant West Texas wind and rising regional demand driven by oil and gas electrification and data centre growth. Fully disentangling these effects, however, is beyond the scope of this study.

Turning to hub price differentials, the Houston hub plays a central role in propagating congestion costs across the system, particularly during cooler periods. $\Delta_{\text{Houston}}$ causally increases $\Delta_{\text{North}}$ and $\Delta_{\text{South}}$ in 6 of 10 cool-period windows (Fig. \ref{figure6}). Beginning in 2020, these effects extend to the West region, either directly from Houston or indirectly through the North hub. This is consistent with Houston's position as an import-reliant demand centre with limited and costly local generation \cite{FellRoh2019}. During warm months $\Delta_{\text{Houston}}$ and $\Delta_{\text{South}}$ remain causally linked across all study windows. The direction of effects varies during warm peak hours, but during warm off-peak hours $\Delta_{\text{South}}$ predominantly drives $\Delta_{\text{Houston}}$, with the effect strengthening over time.

\begin{landscape}
\begin{figure}[htbp]
\centering
\vspace*{-0.5cm}

\newcommand{\colw}{0.13\linewidth}
\newcommand{\headersize}{\scriptsize}
\newcommand{\rowlabelsize}{\scriptsize}

\setlength{\tabcolsep}{1pt}
\renewcommand{\arraystretch}{0.3}

\begin{tabular}{c ccccc}
 & \headersize 2019--2020 & \headersize 2020--2021 & \headersize 2021--2022 & \headersize 2022--2023 & \headersize 2023--2024 \\[2pt]

\raisebox{.9cm}{\rotatebox{90}{\rowlabelsize cool peak}} &
\includegraphics[width=\colw]{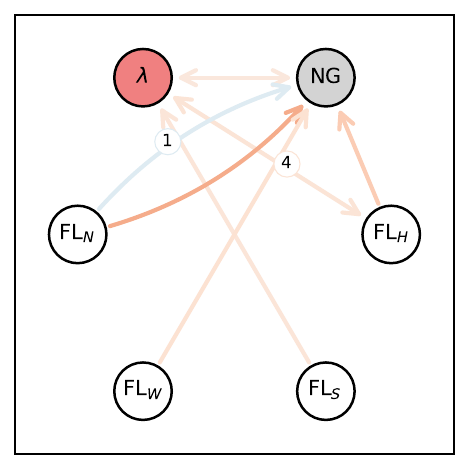} &
\includegraphics[width=\colw]{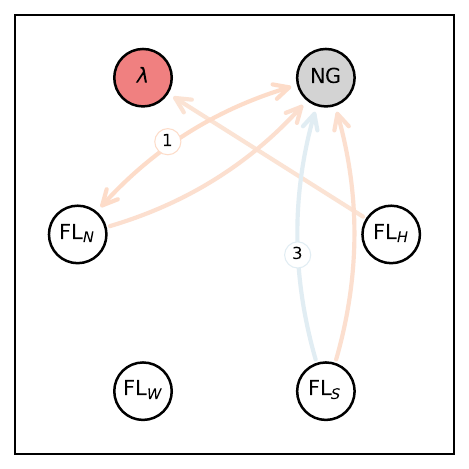} &
\includegraphics[width=\colw]{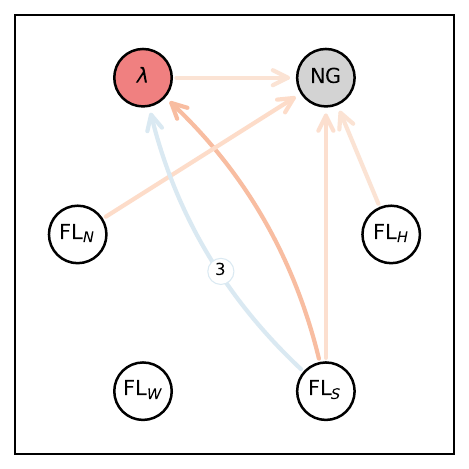} &
\includegraphics[width=\colw]{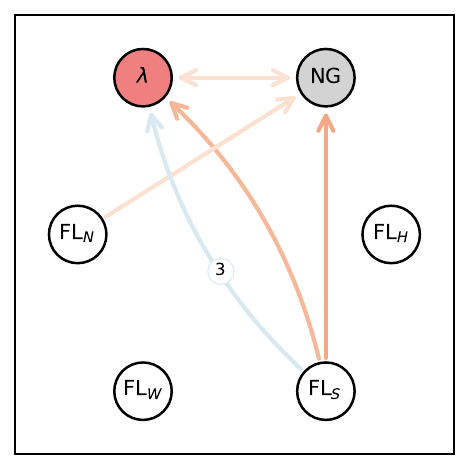} &
\includegraphics[width=\colw]{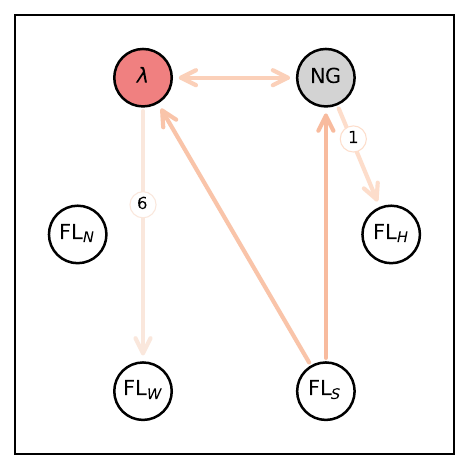} \\

\raisebox{.8cm}{\rotatebox{90}{\rowlabelsize cool off-peak}} &
\includegraphics[width=\colw]{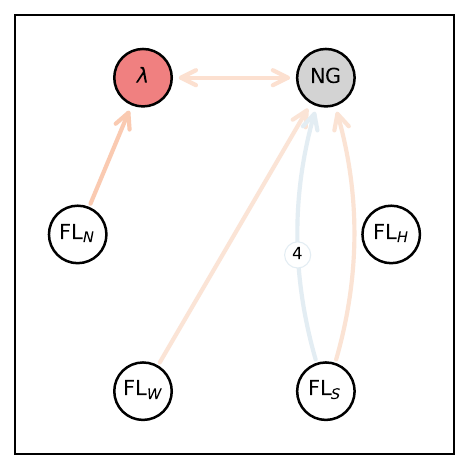} &
\includegraphics[width=\colw]{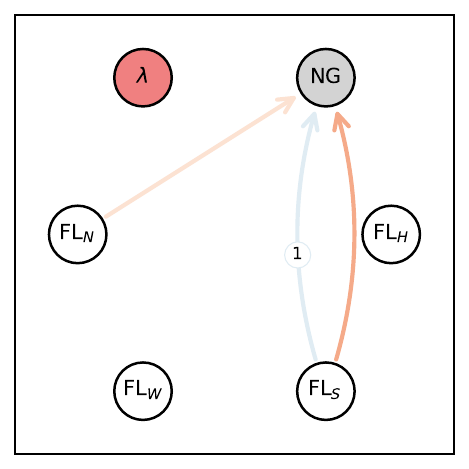} &
\includegraphics[width=\colw]{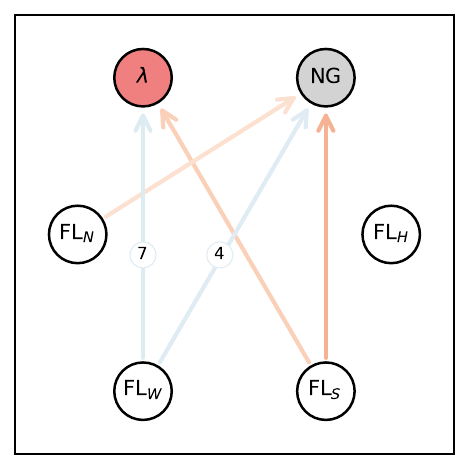} &
\includegraphics[width=\colw]{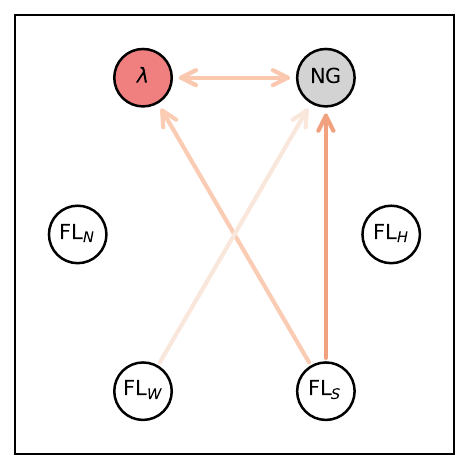} &
\includegraphics[width=\colw]{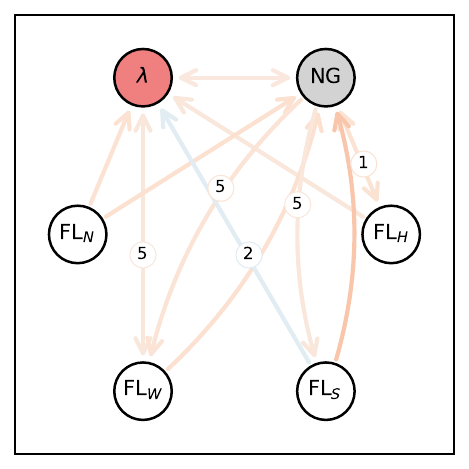} \\

\raisebox{.9cm}{\rotatebox{90}{\rowlabelsize warm peak}} &
\includegraphics[width=\colw]{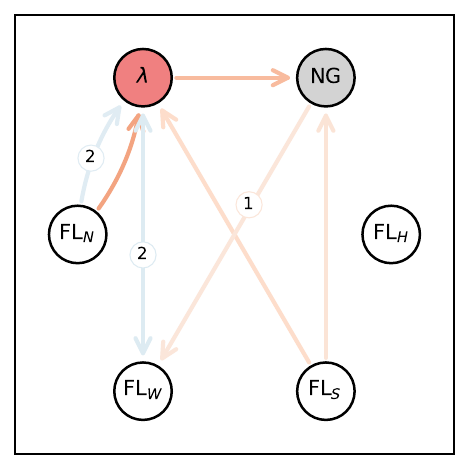} &
\includegraphics[width=\colw]{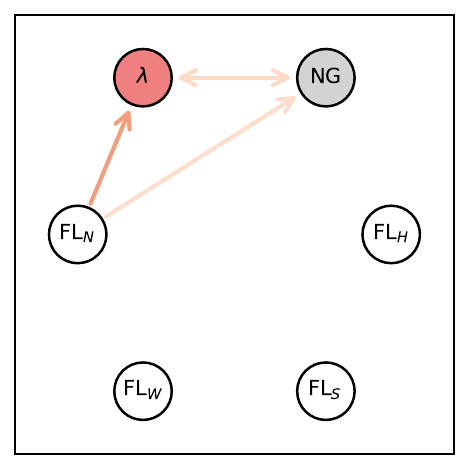} &
\includegraphics[width=\colw]{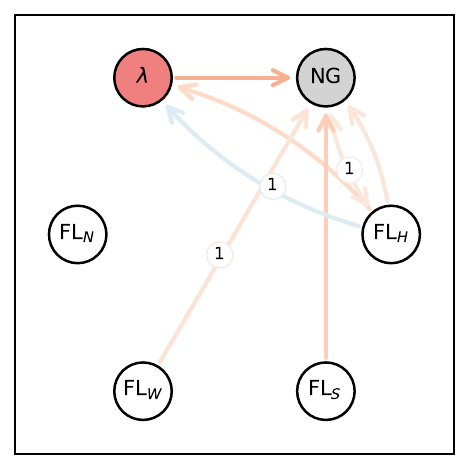} &
\includegraphics[width=\colw]{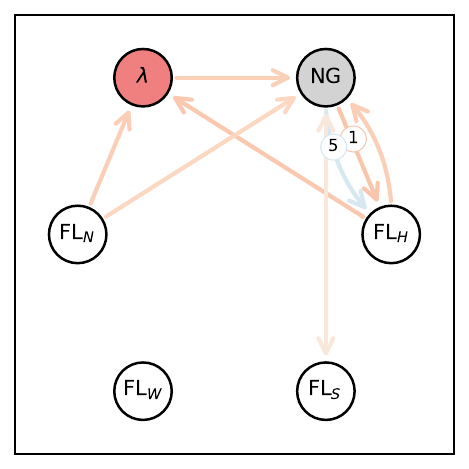} &
\includegraphics[width=\colw]{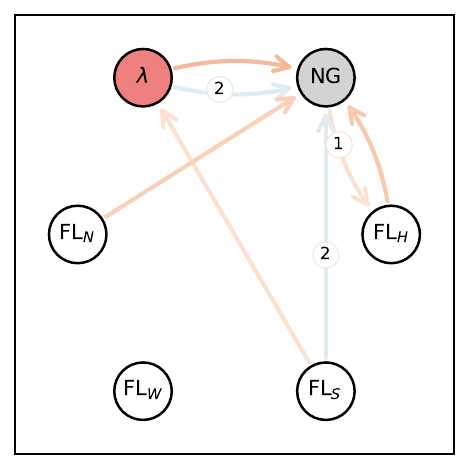} \\

\raisebox{.7cm}{\rotatebox{90}{\rowlabelsize warm off-peak}} &
\includegraphics[width=\colw]{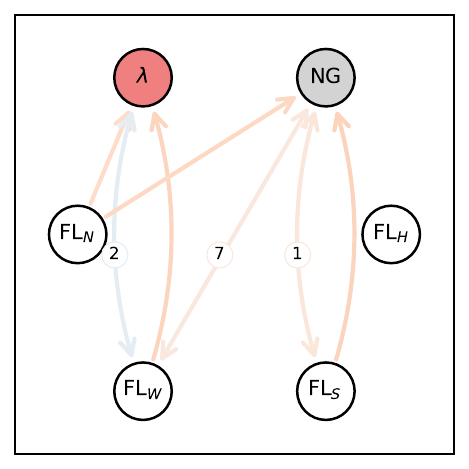} &
\includegraphics[width=\colw]{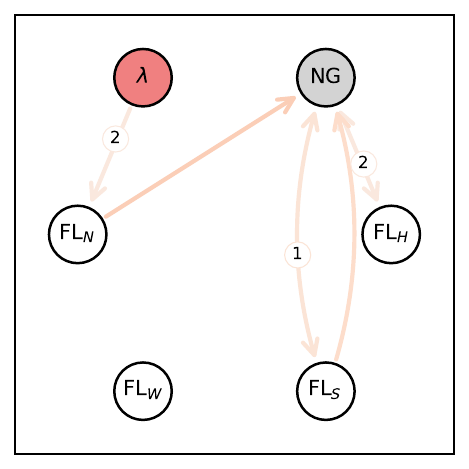} &
\includegraphics[width=\colw]{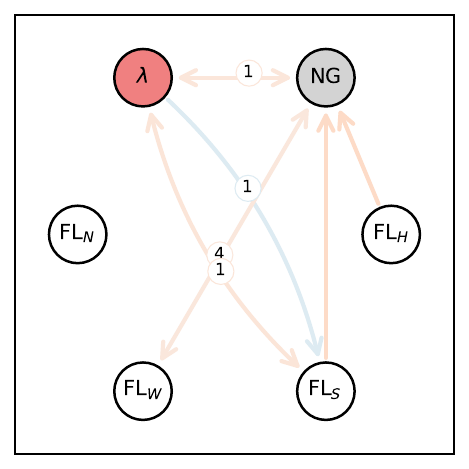} &
\includegraphics[width=\colw]{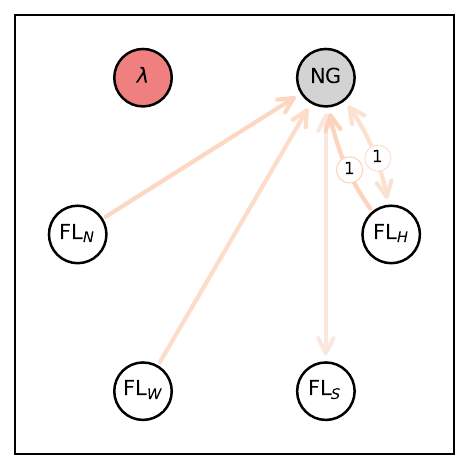} &
\includegraphics[width=\colw]{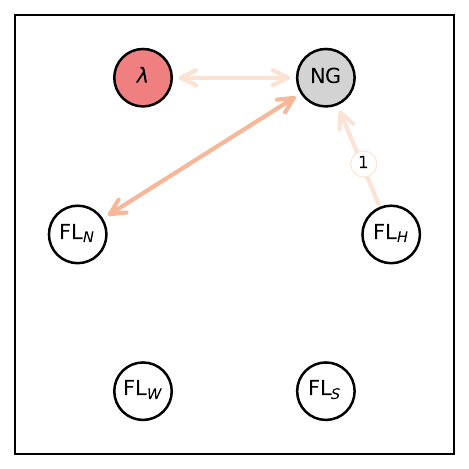} \\[6pt]
 & \multicolumn{3}{l}{\includegraphics[width=0.21\linewidth]{figs/colorbar.pdf}} \\

\end{tabular}

    \caption{Causal effects of regional electricity load on system lambda and natural gas generation. FL$_N$, FL$_S$, FL$_W$ and FL$_H$ denote forecast load in the North, South, West and Houston regions, respectively. Edges between load variables \textbf{are omitted for clarity}, as are autoregressive edges (self-loops). Panels are arranged by market regime (peak/off-peak, warm/cool) over successive two-year windows from 2019 to 2024. Edge colour indicates the sign and magnitude of the causal effect (see colour scale). Edges without a lag label denote contemporaneous relationships (lag 0); numbered edges indicate the lag in days.}
    \label{figure4}
\end{figure}

\begin{figure}[htbp]
\centering
\vspace*{-0.5cm}

\newcommand{\colw}{0.13\linewidth}
\newcommand{\headersize}{\scriptsize}
\newcommand{\rowlabelsize}{\scriptsize}

\setlength{\tabcolsep}{1pt}
\renewcommand{\arraystretch}{0.3}

\begin{tabular}{c ccccc}
 & \headersize 2019--2020 & \headersize 2020--2021 & \headersize 2021--2022 & \headersize 2022--2023 & \headersize 2023--2024 \\[2pt]

\raisebox{.9cm}{\rotatebox{90}{\rowlabelsize cool peak}} &
\includegraphics[width=\colw]{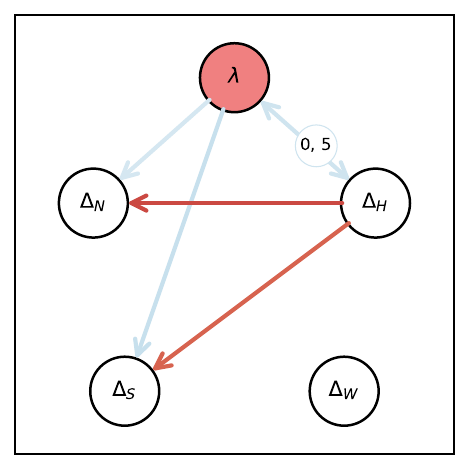} &
\includegraphics[width=\colw]{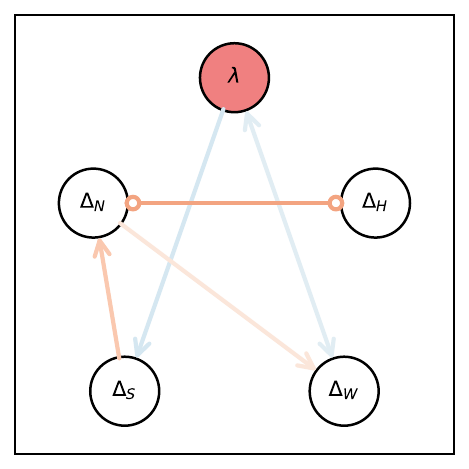} &
\includegraphics[width=\colw]{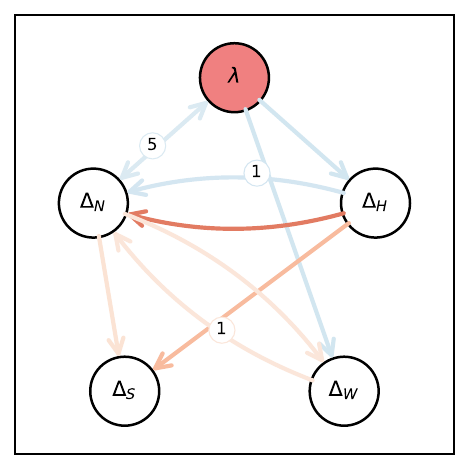} &
\includegraphics[width=\colw]{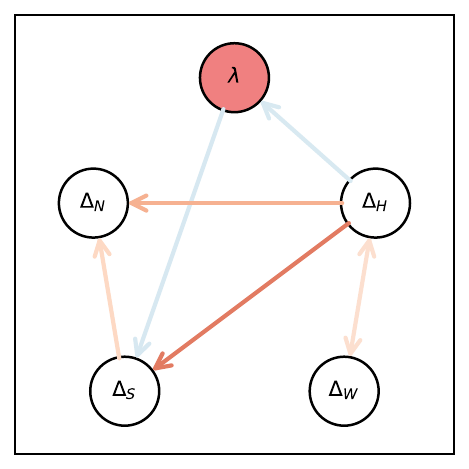} &
\includegraphics[width=\colw]{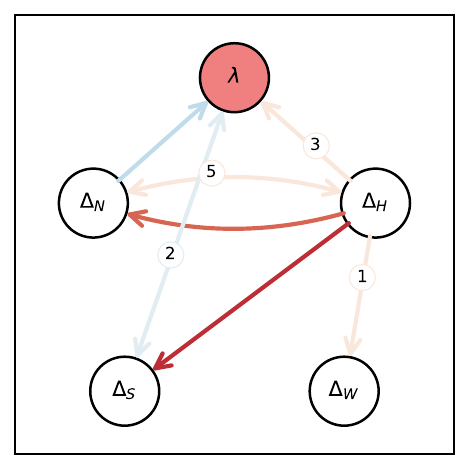} \\

\raisebox{.8cm}{\rotatebox{90}{\rowlabelsize cool off-peak}} &
\includegraphics[width=\colw]{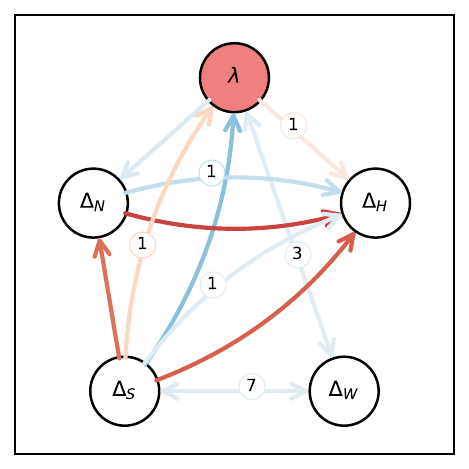} &
\includegraphics[width=\colw]{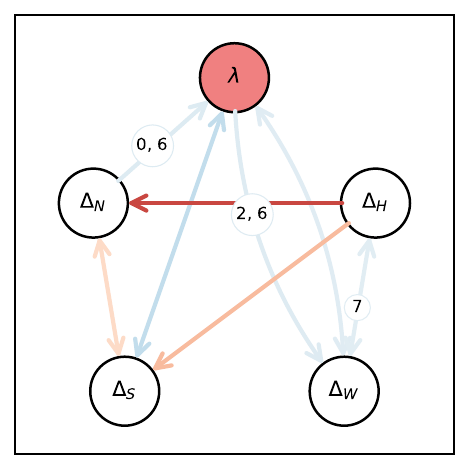} &
\includegraphics[width=\colw]{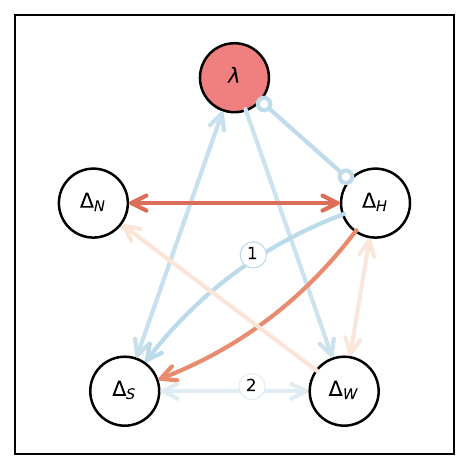} &
\includegraphics[width=\colw]{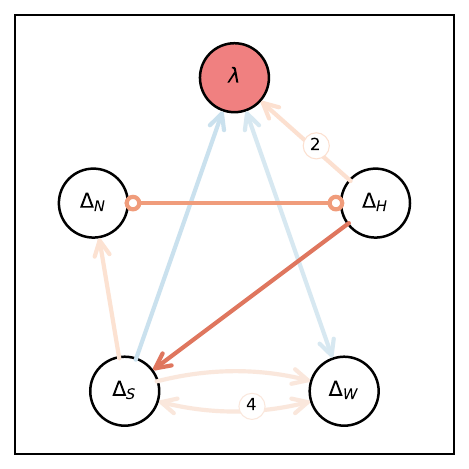} &
\includegraphics[width=\colw]{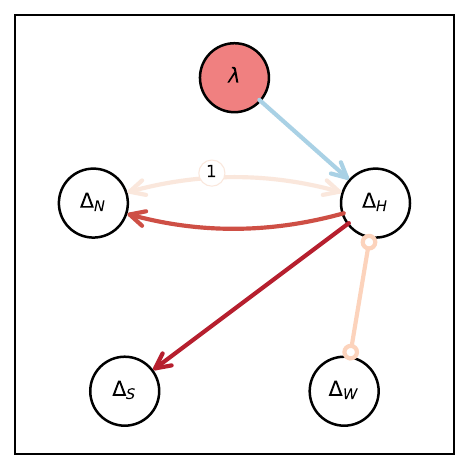} \\

\raisebox{.9cm}{\rotatebox{90}{\rowlabelsize warm peak}} &
\includegraphics[width=\colw]{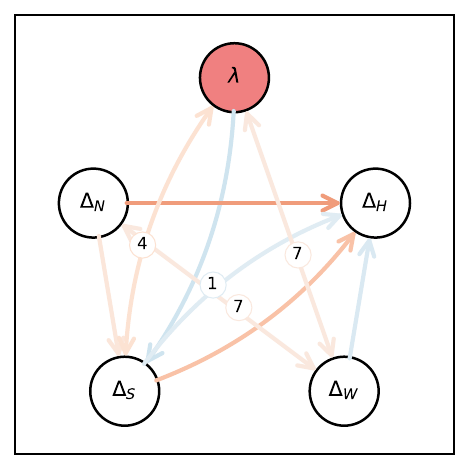} &
\includegraphics[width=\colw]{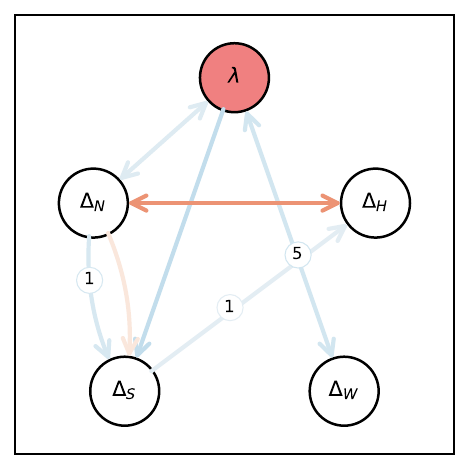} &
\includegraphics[width=\colw]{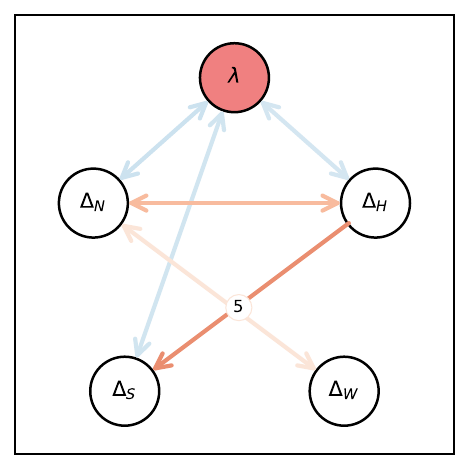} &
\includegraphics[width=\colw]{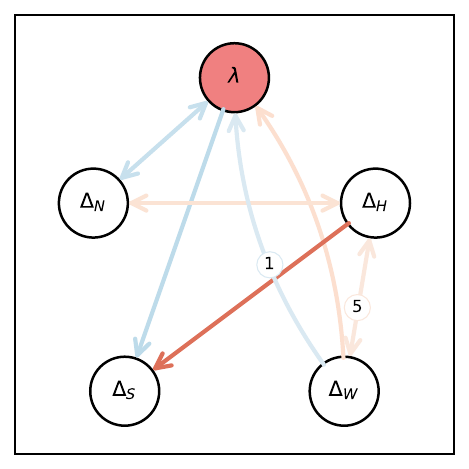} &
\includegraphics[width=\colw]{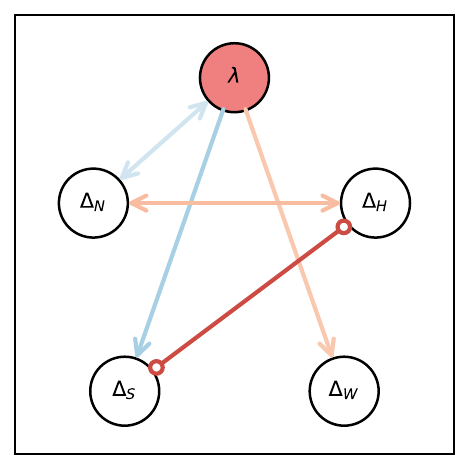} \\

\raisebox{.7cm}{\rotatebox{90}{\rowlabelsize warm off-peak}} &
\includegraphics[width=\colw]{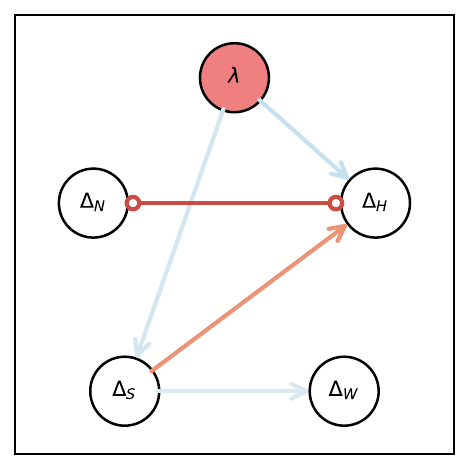} &
\includegraphics[width=\colw]{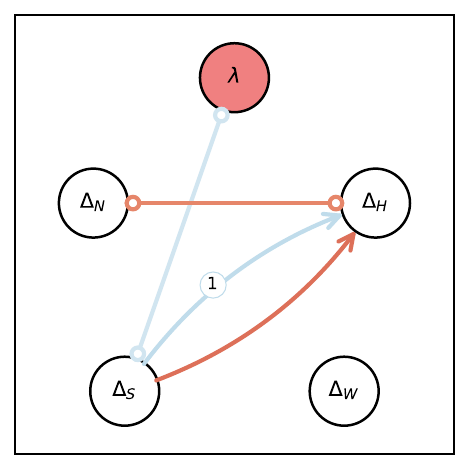} &
\includegraphics[width=\colw]{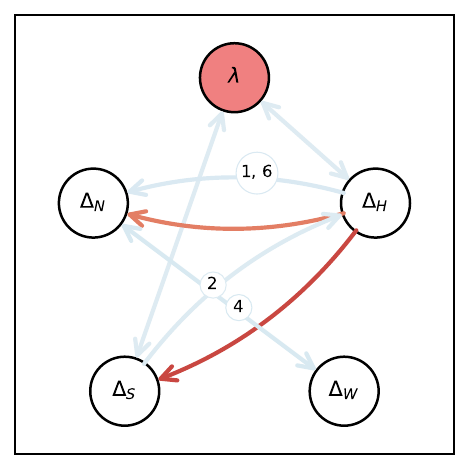} &
\includegraphics[width=\colw]{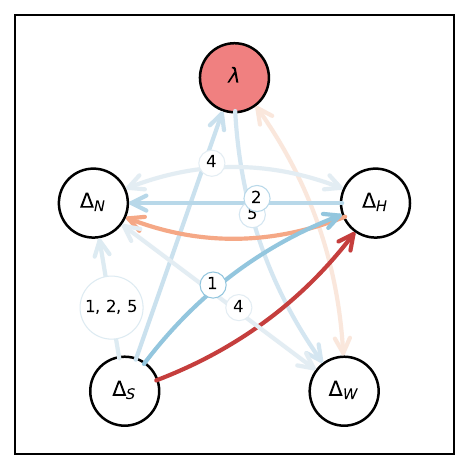} &
\includegraphics[width=\colw]{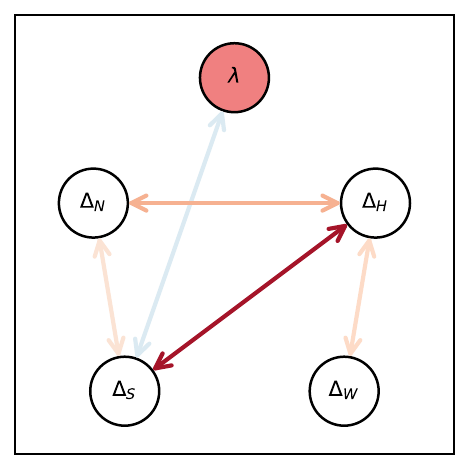} \\[6pt]
 & \multicolumn{3}{l}{\includegraphics[width=0.21\linewidth]{figs/colorbar.pdf}} \\

\end{tabular}

    \caption{Causal relationships between regional price differentials and system lambda. $\Delta_N$, $\Delta_S$, $\Delta_W$, and $\Delta_H$ denote the congestion component of day-ahead prices in the North, South, West and Houston price hubs, respectively. Panels are arranged by market regime (peak/off-peak, warm/cool) over successive two-year windows from 2019 to 2024. Edge colour indicates the sign and magnitude of the causal effect (see colour scale). Edges without a lag label denote contemporaneous relationships (lag 0); numbered edges indicate the lag in days. Autoregressive edges (self-loops) \textbf{are omitted for clarity.}}
    \label{figure6}
\end{figure}
\end{landscape}

\subsection*{Natural gas markets}

Natural gas prices have long been the primary driver of electricity prices in thermal-dominated systems. We examine whether this relationship persists as renewable penetration increases and whether the relative causal importance of different gas hubs has shifted. In Texas, this question is complicated by regional gas hub dynamics and evolving demand patterns. West Texas's Waha hub, located near Permian Basin production, has historically traded at a significant discount to Henry Hub owing to pipeline capacity constraints \cite{eia_ng_2023}. Meanwhile, growing liquefied natural gas (LNG) exports from Houston and ongoing electrification of oil and gas operations in West Texas have reshaped NG demand across the state. 

\begin{landscape}
\begin{figure}[htbp]
\centering
\vspace*{-0.5cm}

\newcommand{\colw}{0.13\linewidth}
\newcommand{\headersize}{\scriptsize}
\newcommand{\rowlabelsize}{\scriptsize}

\setlength{\tabcolsep}{1pt}
\renewcommand{\arraystretch}{0.3}

\begin{tabular}{c ccccc}
 & \headersize 2019--2020 & \headersize 2020--2021 & \headersize 2021--2022 & \headersize 2022--2023 & \headersize 2023--2024 \\[2pt]

\raisebox{.9cm}{\rotatebox{90}{\rowlabelsize cool peak}} &
\includegraphics[width=\colw]{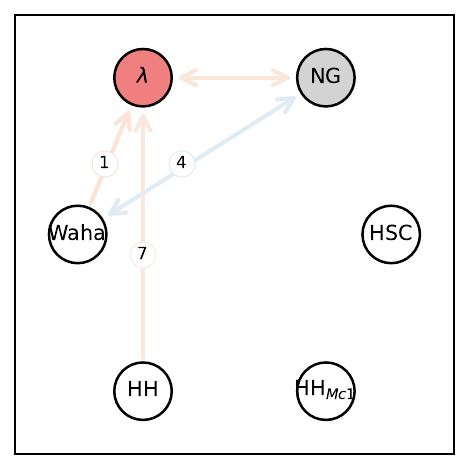} &
\includegraphics[width=\colw]{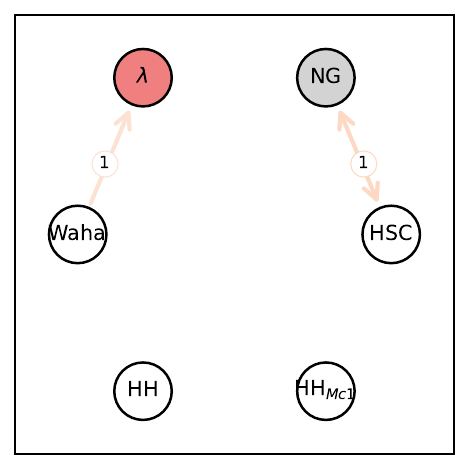} &
\includegraphics[width=\colw]{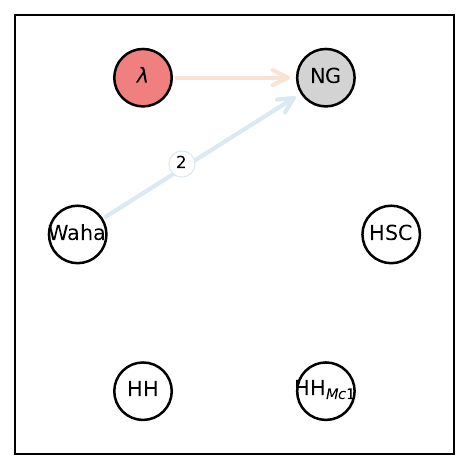} &
\includegraphics[width=\colw]{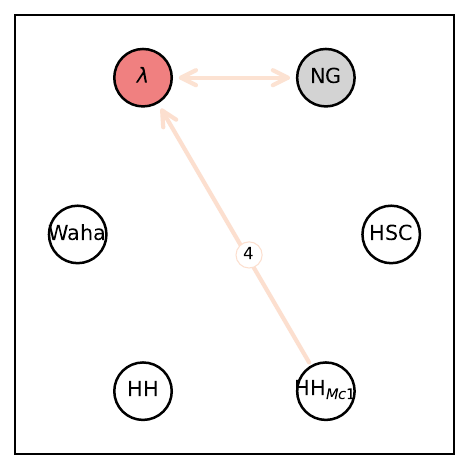} &
\includegraphics[width=\colw]{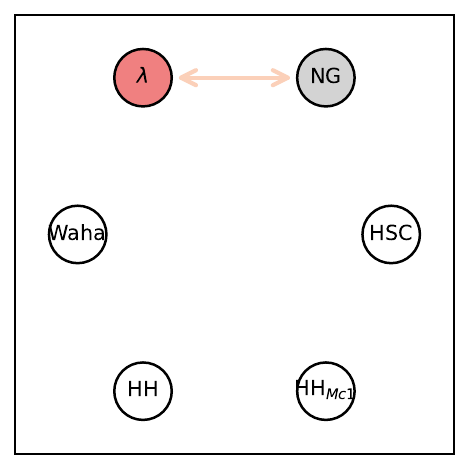} \\

\raisebox{.8cm}{\rotatebox{90}{\rowlabelsize cool off-peak}} &
\includegraphics[width=\colw]{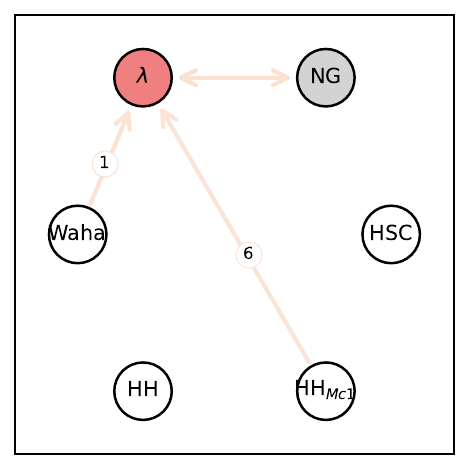} &
\includegraphics[width=\colw]{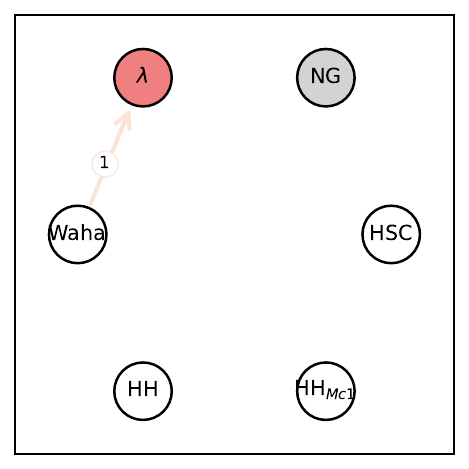} &
\includegraphics[width=\colw]{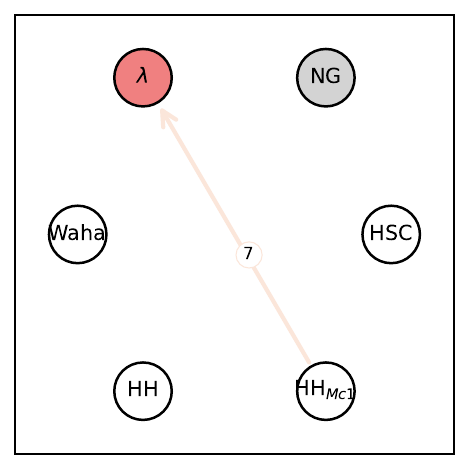} &
\includegraphics[width=\colw]{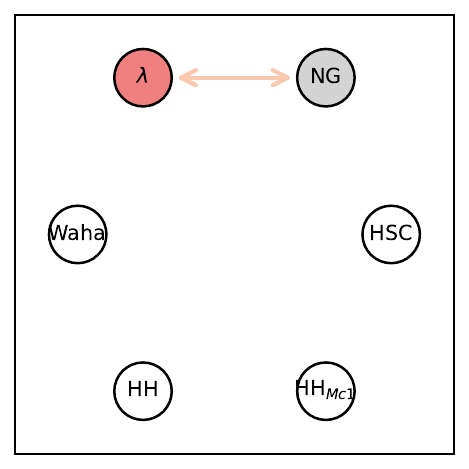} &
\includegraphics[width=\colw]{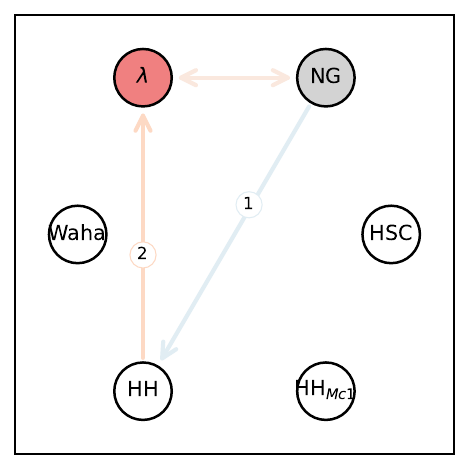} \\

\raisebox{.9cm}{\rotatebox{90}{\rowlabelsize warm peak}} &
\includegraphics[width=\colw]{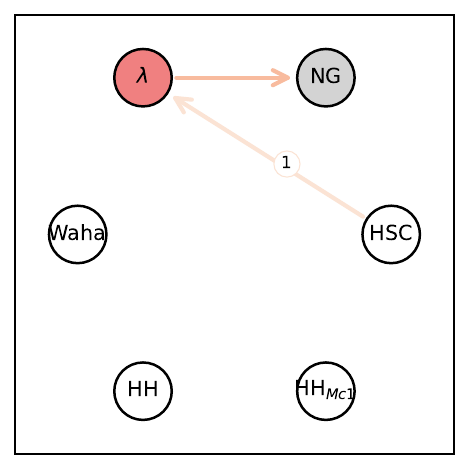} &
\includegraphics[width=\colw]{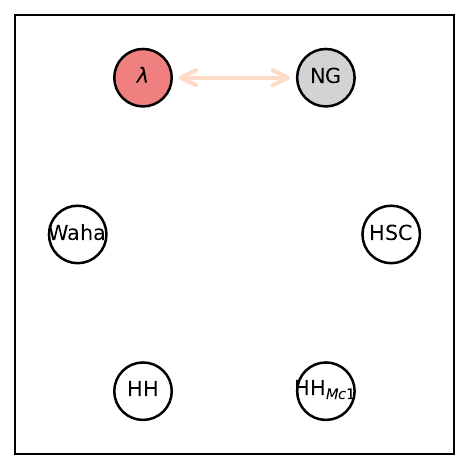} &
\includegraphics[width=\colw]{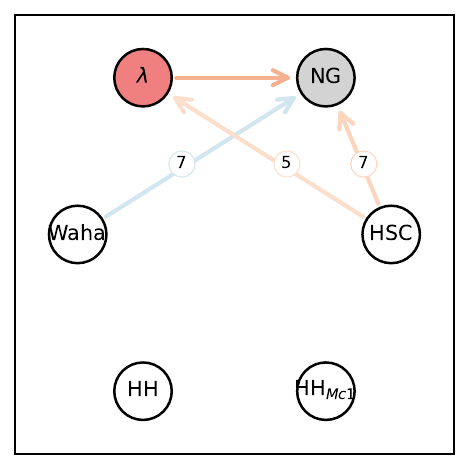} &
\includegraphics[width=\colw]{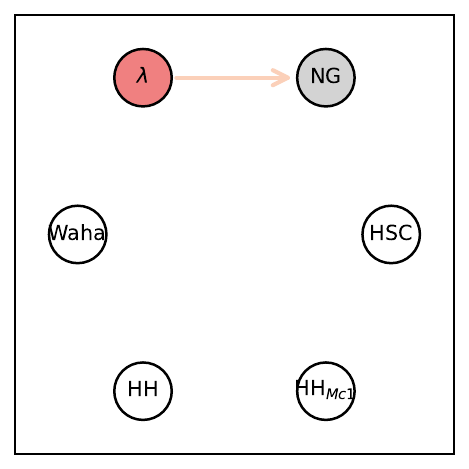} &
\includegraphics[width=\colw]{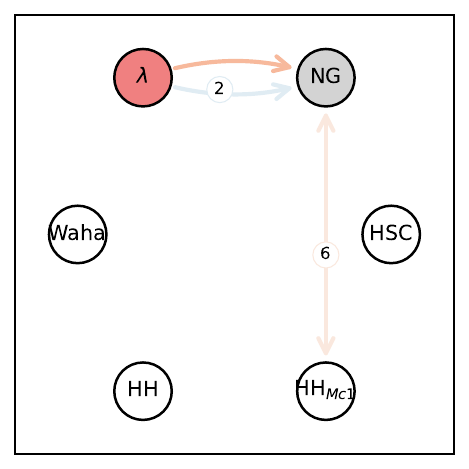} \\

\raisebox{.7cm}{\rotatebox{90}{\rowlabelsize warm off-peak}} &
\includegraphics[width=\colw]{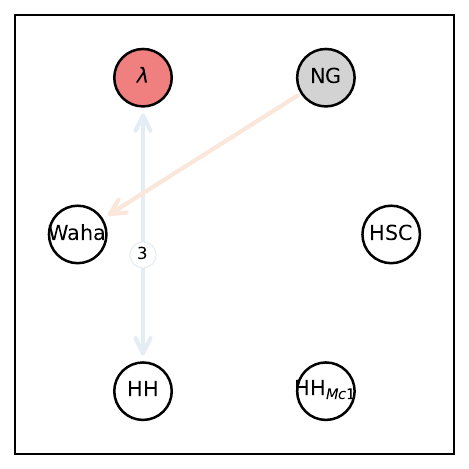} &
\includegraphics[width=\colw]{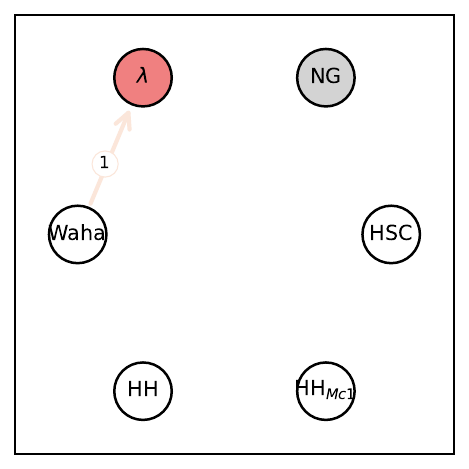} &
\includegraphics[width=\colw]{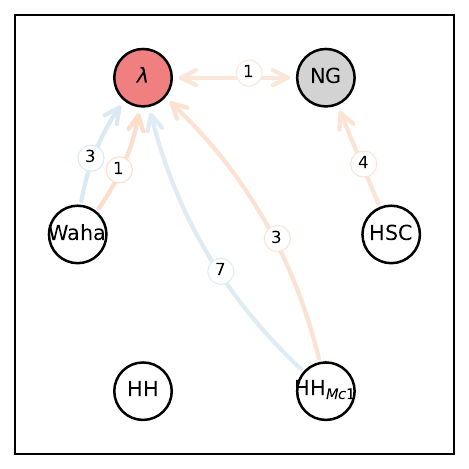} &
\includegraphics[width=\colw]{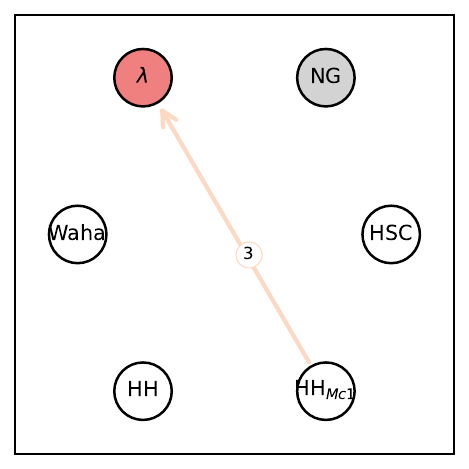} &
\includegraphics[width=\colw]{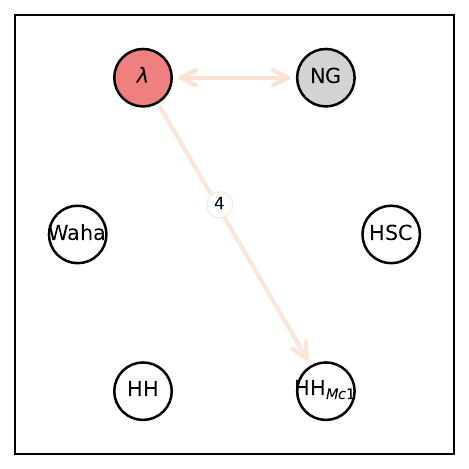} \\[6pt]
 & \multicolumn{3}{l}{\includegraphics[width=0.21\linewidth]{figs/colorbar.pdf}} \\

\end{tabular}

    \caption{Causal effects of natural gas hub prices on system lambda and natural gas generation. HH, HH$_{\text{Mc1}}$, and HSC refer to Henry Hub day-ahead, Henry Hub front-month futures, and Houston Ship Channel day-ahead prices, respectively (see Methods, Table 1). Panels are arranged by market regime (peak/off-peak, warm/cool) over successive two-year windows from 2019 to 2024. Edge colour indicates the sign and magnitude of the causal effect (see colour scale). Edges without a lag label denote contemporaneous relationships (lag 0); numbered edges indicate the lag in days. Autoregressive edges (self-loops) \textbf{are omitted for clarity.}}
    \label{figure5}
\end{figure}
\end{landscape}

Natural gas can affect system lambda through two channels: fuel prices and the quantity of gas-fired generation, with the latter itself causally influenced by regional load and renewable generation (as shown in previous subsections). Our analysis indicates that natural gas prices continue to exert a statistically significant causal effect on system lambda through 2024, but the link appears intermittently and is small in magnitude compared to wind generation effects, which are 3.4 times larger on average. As Fig. \ref{figure5} shows, NG prices drive system lambda in both warm and cool periods throughout 2019--2024, with no evidence that the relationship fully disappears despite substantial renewable growth. 

The relevant natural gas hubs, however, change markedly over the period. During 2019–2021, Waha prices play a dominant role in driving system lambda, particularly during cooler months (Fig. \ref{figure5}). As Waha and Henry Hub prices converge following pipeline expansion in 2021 \cite{eia_ng_2023}, Waha's direct causal influence on system lambda diminishes in our causal graphs. Post-2021, Henry Hub day-ahead and futures prices emerge as the primary gas price drivers of system lambda in both warm and cool periods whenever a link appears, though with similarly small effect sizes (Fig. \ref{figure5}). This change suggests that as local infrastructure constraints in West Texas eased, electricity prices became more responsive to national natural gas market dynamics. 

Gas prices show little direct causal influence on NG generation volumes, doing so in only 3 of 20 periods. The relationship between NG generation and system lambda is predominantly bidirectional (9 of 20 periods), reflecting shared underlying drivers. During warm peak hours, however, a directed edge from lambda to NG generation appears, in line with higher day-ahead clearing prices under tight system conditions driving greater NG-fired dispatch in real time. By contrast, we find only sporadic causal links between regional price differentials and either natural gas prices or generation volumes, with no salient pattern appearing across or within the periods analysed. This suggests that congestion costs are primarily driven by the spatial mismatch between renewable generation and load rather than fuel price dynamics.

\section*{Discussion}\label{sec6}

In this study, we show that forecast wind generation from West Texas has become the dominant causal driver of day-ahead system lambda (the system-wide energy component of prices) in ERCOT. West wind exerts causal effects more than three times greater than natural gas prices, challenging the conventional view of the Texas power market as gas price-driven. This reordering in causal importance points to a broader transformation in electricity price formation, as renewable integration and evolving demand patterns lead to a rebalancing of supply-demand forces. It also implies that shocks to wind availability will affect electricity prices more strongly than do fluctuations in natural gas prices, exposing the power market to greater weather-dependent risk. Such exposure is likely to intensify as climate change alters wind patterns and increases summer heat and winter cold extremes \cite{losada2020potential}. 

Yet, wind's price-dampening effects are not uniform. Recent work highlights an insurance value of renewables in reducing electricity price exposure to fossil fuel shocks, though with significant variation across hours of the day \cite{navia_simon_power_2025}. Our findings extend this observation by showing that wind's price-dampening effects are both temporally and seasonally asymmetric, weakening during the system's most stressed periods. During warm peak hours, the effect has roughly halved over our study period, despite continued wind capacity growth and no significant rise in curtailment (Supplementary Fig. \ref{fig:figureS2}). This suggests that wind and transmission expansion may not have kept pace with demand growth during constrained periods. For wind generators, this weakening translates into reduced self-cannibalization during high-price hours \cite{navia_simon_power_2025, prol2020cannibalization}. For the system, however, it erodes price relief precisely when it is most needed. For policymakers, this points to the need for better alignment of capacity additions with demand growth during peak stress hours. In the near term, incentivizing storage deployment or temporal flexibility in large loads could compensate for the slower rate of capacity expansion and more effectively utilize existing grid capacity \cite{norris_load_2025}.

Wind generation also reshapes hub price differentials across ERCOT's regions. Both West and non-West wind alleviate price differentials locally, but West Texas wind redistributes them elsewhere in seasonally shifting patterns. During warm peak hours, Wind$_{west}$ increases the North Texas differential, indicating transmission constraints on west-to-north power flows. Expanding transmission along this corridor and within constrained regions would help address this issue. During cool peak hours, the spillover pattern shifts to Houston instead, as reduced gas dispatch in North Texas forces Houston onto costlier local units \cite{FellRoh2019}. Unlike the warm-season constraint, the latter cannot be addressed through transmission expansion alone and requires flexible resources or storage within Houston. Since Houston consistently acts as a source of price differential spillovers to other regions, particularly during cooler periods, dampening its price differentials would have the added benefit of reducing congestion cost propagation across the network. 

The null result for solar generation should be interpreted with caution, as our peak-hour definition includes hours with minimal solar output. As solar penetration in Texas continues to grow rapidly, its causal role in price formation is likely to become more detectable. Solar generation already displaces natural gas generation during midday hours \cite{eia2025ercot}, and may also counterbalance the warm-season weakening of wind's causal effects in those hours. Examining solar's causal effects at finer temporal resolution is an important direction for future work, and the same causal framework can be readily applied.

Despite the rapid growth of renewables, natural gas prices continue to causally affect system lambda throughout our study period, though intermittently and with modest magnitude relative to wind. Notably, the gas hub that causally drives system lambda has shifted: during 2019–2021, Waha hub prices drove system lambda, but following the expansion of pipeline capacity from West to Eastern and coastal Texas and the resulting convergence of regional gas prices, this influence migrated to Henry Hub. Texas electricity prices are now more responsive to national gas market fundamentals than to regional supply conditions. For market participants managing price risk, this argues for tracking Henry Hub rather than Waha as the relevant gas benchmark, though a reversal is plausible if renewed growth in Permian Basin oil production re-creates pipeline bottlenecks and restores the east-west gas price spread.

The causal structure of price formation is also shifting on the demand side. During cooler periods, South Texas load emerges as a driver of both natural gas generation and system lambda from 2021 onward, as rising local demand and binding transmission constraints limit the flow of non-west wind to other regions. Expanding transmission capacity within and out of the South zone could help alleviate this constraint, as suggested by the reduction in congestion following recent transmission investments in this region \cite{potomaceconomics2025}. West Texas load follows a distinct trajectory: it disappears as a causal driver of natural gas generation and/or system lambda during 2020–2021 before re-emerging, predominantly in off-peak hours. Electricity demand in the West zone continues to grow, driven by oil and gas electrification and co-location of data centres with renewable generation in the region. If this trend continues, West Texas may increasingly consume its own wind output, reducing exports and reshaping system-wide price dynamics.

Our findings, across wind's changing price effects, the shift of the dominant natural gas price hub, and the evolving dynamics of South and West Texas load,  demonstrate that the causal structure of electricity price formation is not static. Analyses that assume fixed relationships risk missing these shifts and mischaracterising how prices form. The framework presented here complements energy system modelling and forecasting approaches by tracking how causal relationships evolve over time, revealing which drivers remain stable, which vary with conditions such as season or demand levels, and which emerge or disappear as the system transforms. This is particularly relevant given the wide adoption of of data-driven and machine learning tools across the power sector by traders, investors, and grid operators. Our framework provides causally grounded inputs that can improve the robustness of these models to structural change and guard against spurious relationships, such as the confounded association between non-west wind and natural gas generation identified in our study. For regulators, the framework can help assess whether price signals propagate as market design intends and identify where they do not, pointing to potential market inefficiencies.

Although we apply the framework to Texas, the core findings carry broader relevance for electricity markets worldwide undergoing structural change. The emergence of wind as the dominant causal driver of electricity prices, redistributing congestion costs away from renewable-rich regions, is likely to occur in other markets with growing renewable penetration. Similarly, the shift in the influence of natural gas hubs following pipeline expansion illustrates how energy infrastructure investments can alter the causal structure of electricity price formation. These dynamics have direct implications for how price risks are assessed and managed, and for capacity and infrastructure decisions that depend on identifying which factors causally shape prices; our analysis indicates, for instance, where storage or transmission expansion would most effectively alleviate the congestion (price differentials). The emergence of South and West Texas load as causal drivers further shows how demand-side dynamics increasingly shape market outcomes. As AI workloads expand and large consumers participate in infrastructure decisions through behind-the-meter generation, storage, and long-term contracts, the causal role of the demand side will grow. Within Texas, the causal restructuring will continue as solar and battery storage scale, electrification and large new loads raise demand, and market rules adapt. How investors, planners, and market participants respond to these shifts will determine the trajectory and pace of the transition, along with the affordability, reliability, and carbon intensity of electricity supply.

\clearpage
\section*{Methods}\label{sec:methods}

\subsection*{Data} 

ERCOT has operated a nodal wholesale market since 2010, in which electricity prices vary by location as locational marginal prices (LMPs), reflecting the fact that least-cost generation cannot always serve demand everywhere due to physical transmission constraints. Day-ahead LMPs in ERCOT consist of two components: (i) the system lambda, the shadow price of the system-wide power balance constraint, uniform across all nodes, and (ii) a node-specific congestion component, reflecting the shadow price of binding transmission constraints \cite{ercot_da_ops_2024}. Nodal LMPs are aggregated into trading hubs, whose prices are simple average of constituent bus LMPs, and into load zones, whose prices are load-weighted averages used for settlement of demand. Our analysis focuses on trading hub prices.

Table \ref{table1} lists the variables included in our analysis and their data sources. System lambda, hub electricity prices and wind speed values are averaged over peak or off-peak hours; natural gas hub prices are daily closing values; temperature data are daily averages. The remaining variables are summed over the relevant hours. For natural gas hub prices, missing values are forward-filled using the last available price. For average temperature, missing values are first filled using the mean of daily minimum and maximum temperatures from NOAA; five remaining missing dates are filled by interpolation. Missing forecast load data are excluded from the analysis using LPCMCI's masking function (see model setup below) rather than imputed, as forecast load is closely related to system conditions and cannot be reliably imputed from adjacent observations. The weather variables are principal components derived from average temperatures, total global horizontal irradiance (GHI) and average wind speeds across 12 weather stations (Table \ref{table1}). We retain five components following the Kaiser–Guttman criterion, which together explain 87\% of the variance. This reduces the dimensionality of the weather variables while capturing broad spatial patterns such as weather fronts across Texas.

We distinguish between wind generation in West Texas and the rest of the state, due to the markedly different generation profiles and geographic concentration of capacity in these regions \cite{slusarewicz2018assessing}. Consistent with our focus on the DAM, we use forecasted rather than realised renewable generation and regional demand, as these forecasts shape market expectations and inform bidding behaviour in the DAM.

\setcounter{table}{0}
\renewcommand{\thetable}{\arabic{table}}

\begin{table}[ht]
\small
\caption{Study variables and data sources}\label{table1}
\renewcommand{\arraystretch}{1.2}
\setlength{\extrarowheight}{0pt}
\begin{tabular}{@{}p{0.3cm}L{3cm}L{5.3cm}p{1.5cm}L{4.2cm}@{}}
\toprule
& \textbf{Variable} & \textbf{Description} & \textbf{Source} & \textbf{Details} \\
\midrule

1 & System lambda ($\lambda$) & System-wide day-ahead marginal energy cost & GridStatus \cite{gridstatusio} & \path{"ercot_dam_system_lambda"} \\
\addlinespace
\arrayrulecolor{gray!40}\hline\arrayrulecolor{black}
2 & Hub price differentials ($\Delta_h$) & Day-ahead hub $h$ price minus system lambda (variable 1) & ERCOT \cite{ercot_data} & \path{"NP4-180-ER"}\\
\addlinespace
\arrayrulecolor{gray!40}\hline\arrayrulecolor{black}
3 & Wind forecast (West) & Short-term wind power forecast, load zone West (\textnormal{\small stwpf\_lz\_west)} & GridStatus \cite{gridstatusio}& \path{"ercot_wind_actual_and_forecast_hourly"}\\
\addlinespace
\arrayrulecolor{gray!40}\hline\arrayrulecolor{black}
4 & Wind forecast (non-West) & System-wide short-term wind power forecast minus West Texas (\textnormal{\small stwpf\_lz\_system\_wide}, \textnormal{\small stwpf\_lz\_west}) & GridStatus \cite{gridstatusio}
& \path{"ercot_wind_actual_and_forecast_hourly"}\\
\addlinespace
\arrayrulecolor{gray!40}\hline\arrayrulecolor{black}
5 & Solar forecast & System-wide short-term solar power forecast (\textnormal{\small stppf\_system\_wide}) & GridStatus \cite{gridstatusio} & \path{"ercot_solar_actual_and_forecast_hourly"}\\
\addlinespace
\arrayrulecolor{gray!40}\hline\arrayrulecolor{black}
6 & Natural gas generation & Total generation volume from natural gas & EIA \cite{eia_gridmonitor}& \textnormal{\small form 930; region ERCO} \\
\addlinespace
\arrayrulecolor{gray!40}\hline\arrayrulecolor{black}
7 & Natural gas prices & \makecell[tl]{Waha hub (day-ahead) \\ Houston Ship Channel (day-ahead) \\ Henry Hub (day-ahead) \\ Henry Hub (front-month futures)} & LSEG (Refinitiv) \cite{Refinitiv}
& \makecell[tl] {\textnormal{\small NG-WAH-WTX-SNL} \\ \textnormal{\small NG-PHSC-TX-SNL}\\ \textnormal{\small NG-W-HH-SNL}\\ \textnormal{\small NGc1}}\\
\addlinespace
\arrayrulecolor{gray!40}\hline\arrayrulecolor{black}
8 & Load forecasts & Forecast load by ERCOT load zone (North, West, South, Houston) & GridStatus \cite{gridstatusio} & \path{"ercot_load_forecast_by_forecast_zone"}\\
\addlinespace
\arrayrulecolor{gray!40}\hline\arrayrulecolor{black}
9 & Weather variables & Principal components derived from average temperature, wind speed, and global horizontal irradiance (GHI) across 12 Texas weather stations\tnote{*} & NOAA \cite{menne2012overview, ghcnd}; NLR \cite{sengupta2018national} & \makecell[tl]{GHCN (temperature) \\ NSRDB (wind speed,\\ GHI)} \\
\addlinespace
\arrayrulecolor{gray!40}\hline\arrayrulecolor{black}
10 & Controls & \makecell[tl]{Seasonal harmonics (annual, semi-annual,\\
weekly) \\ Linear trend \\ Weekend indicator} & & Derived \\

\bottomrule
\end{tabular}
\begin{tablenotes}
\small
\item[*] Locations include Abilene, Austin, Brownsville, Corpus Christi, Dallas, Houston, Lubbock, McAllen, Midland, San Antonio, Tyler, Wichita Falls.
\end{tablenotes}
\end{table}

\clearpage
\subsection*{Causal discovery approach}

\paragraph*{Comparison to Granger causality} 
For a long time, Granger causality \cite{granger1969investigating, granger1980testing} was the standard approach for analyzing causality in time series data. It holds that $X$ Granger causes $Y$ whenever past values of $X$ help predict $Y$ from its own past, conditioned on all other variables in the system \cite{mastakouri2021necessary}. For example, wind energy output would Granger cause natural gas generation, if knowing the wind output from yesterday and the days before helps predict the natural gas generation volume of today when the influence of \emph{all} other relevant factors, such as historical gas prices and demand fluctuations, is accounted for. Despite the wide use of Granger causality, studies have highlighted the limitations of this approach when its strict assumptions are violated \cite{peters2017elements}. First and foremost, Granger causality assumes that \emph{all} variables are observed, an assumption that is almost never met in practice. Moreover, if there is a high autocorrelation, in our example in the variable $Y$, then Granger causality typically has low detection power; see the discussion in \cite{runge2019detecting}. Finally, Granger causality suffers from the curse of dimensionality even for time series with relatively few components; see also \cite{runge2019detecting}.

\paragraph*{The LPCMCI algorithm}
This study builds on the LPCMCI algorithm, a state-of-the-art method for discovering causal relationships from observational time series data under minimal assumptions \cite{gerhardus2020high-recall}. 
In recent decades, substantial progress has been made in developing rigorous methods for causal discovery \cite{pearl2009causality, spirtes2000causation}. They are largely based on the notion of structural causal models, which we briefly introduce.

Consider a multi-variate time series $X=(X_t^i)$ where $t=0, \ldots T$ is the time index and $i=0, \ldots, k$ is the index of the component. In our setting, each component is a variable such as wind production in West Texas or the natural gas prices at the Waha hub. The time series is said to follow a structural causal model if each component is a function of its direct causal drivers and an additional variable that models noise. The direct causal drivers of component $i$ at time $t$ are called the \emph{causal parents} and are denoted by $\pa(X_t^i) \subseteq \{X_s^j : j=0, \ldots, k, \text{ and } s=0, \ldots, t\}$. More formally, a structural causal model requires that 
\begin{equation*} \label{def:structural-causal-model}
    X_t^i := f_i(\pa(X_t^i), \varepsilon_t^i),
\end{equation*}
where the noise variables $\varepsilon_t^i$ are jointly independent, and the functions $f_i$ depend non-trivially on their causal parents. If $X_s^j$ is in the set of causal parents of $X_t^i$, then we say that there is a \emph{causal effect} of $X_s^j$ on $X_t^i$. More intuitively, $X_s^j$ has a causal effect on $X_t^i$ if intervening  on $X_s^j$ (i.e.~changing its value) has an effect on the value of $X_t^i$, which is captured through the function $f_i$. 

The goal of causal discovery is to recover the sets of causal parents for each variable. We impose the standard assumption that the time series satisfies \emph{causal stationarity}, that is, edges repeat over time. This requires that $X_{t-s}^{i}$ is a causal parent of $X_t^j$ for $s \geq 0$ if and only if  $X_{t'-s}^{i}$ is a causal parent of $X_{t'}^j$ for all times $t'$. Since we are not aware of any method for causal discovery that does not require causal stationarity, we conduct separate analyses for the warmer and cooler months of the year, and for peak and off-peak hours. 

We adopt two further natural assumptions: (i) causal relationships respect time, that is, future variables are not causal parents of past variables, in formulas $X_{t}^i \not\in \pa(X_{t-s}^j)$ for $s >0$; and (ii) causal relationships are acyclic, i.e., if $X_{t-s}^j \in \pa(X_t^i)$, then $X_t^i \not\in \pa(X_{t}^j)$. 

If there is a causal relationship between two variables with time lag $s \geq 1$, then the relationship is said to be \emph{lagged}, while relationships with time lag $s=0$ are said to be \emph{contemporaneous}. We interpret contemporaneous links as causal effects that appear within the same time point. In our analysis, this corresponds to cause-and-effect relationships that occur simultaneously within the same block of peak (or off-peak) hours on a given day. 

Importantly, we allow for latent (unobserved) variables in our analysis, that is, we assume to only observe a subset of the components of the time series. All unobserved variables that could potentially influence our analysis such as transmission or generator outages correspond to latent components of the time series. In the presence of latent variables we cannot differentiate between \emph{direct effects}, where one variable is the causal parents of the other, or \emph{indirect effects} that are mediated by a latent variable. Moreover, latent variables may act as confounders between two variables. In this case, the association between two observed variables is fully explained by a latent variable and there is no causal effect.

As outlined above, the main advances of LPCMCI are its ability to accommodate nonlinear dependencies $f_i$ and latent variables. Accounting for possible presence of latent variables, the algorithm aims to recover as many causal relationships as possible. 
The algorithm proceeds in two phases. First, it discovers whether a pair of variables $X_{T-s}^{i}$ and $X_T^j$ is conditionally independent given any subset of the other variables. For pairs that are not detected to be conditionally independent, it concludes that there is either a (direct or indirect) causal effect between $X_{T-s}^{i}$ and $X_T^j$, or that there is latent confounding explaining the association.
In the second phase, the algorithm then distinguishes confounding from causal relationships and determines causal direction using time structure, acyclicity, and further technical constraints (for details, see \cite{gerhardus2020high-recall}). LPCMCI assumes causal stationarity and requires a prespecified maximal time lag $\tau \geq 0$. That is, it only detects causal relationships between $X_{T-s}^{i}$ and $X_T^j$ with a maximum lag $s \leq \tau$. Hence, LPCMCI requires two important inputs: A choice of a conditional independence test and the maximal time lag $\tau$ that is considered. 

\paragraph{Conditional independence test}
When employing the LPCMCI algorithm, we use the ``robust partial correlation'' measure for testing conditional independence, which builds on the literature of nonparanormal models \cite{harris2013pc, liu2012highdimensional, liu2009thenonparanormal} and is implemented in \texttt{tigramite}. It is more robust and assumption-lean than ordinary partial correlation, while retaining aspects of the interpretability and computational advantages.

The test is a two-stage procedure to test the null hypothesis $X \indep Y \, | \, \mathbf{Z}$ for two random variables $X,Y$ and a random vector $\mathbf{Z}$. The first step is a monotone transformation that maps the original values of each variable to a standard normal scale. Since the transformation is monotone, it is invariant to the ordering of the values of each variable. The second step is to apply Pearsons correlations test on the transformed variables by accounting for, i.e., regressing out, the variables in $\mathbf{Z}$. That is, we compare the partial correlation coefficient of the transformed variables to students $t$-distribution with $T-|\mathbf{Z}|-2$ degrees of freedom. 

Note that the testing procedure is designed for detecting dependencies that are monotone while allowing the variables to have non-Gaussian distributions. The sign of the partial correlation indicates whether larger values of $X$ tend to be associated with larger or smaller values of $Y$, after adjusting for the variables $\mathbf{Z}$. The magnitude reflects the strength of this conditional association. If the association can be interpreted as a causal effect, i.e., if the edge between $X$ and $Y$ can be oriented to point into one direction, the association can be interpreted causally and the magnitude of the partial correlation coefficient represents the \emph{strength} of the causal effects. %

\paragraph{Algorithm parameters} 
We use the implementation of the LPCMCI algorithm that is contained in version 5.2 of the \texttt{tigramite} Python package. We set $pc_{alpha}$ = 0.05 and choose a maximum time lag of $\tau_{\max}=7$, corresponding to one week, to capture the well recognized weekly seasonality in electricity prices \cite{weron_electricity_2014}. For computational reasons, we restrict the maximum size of conditioning sets considered in conditional independence tests to three when running the LPCMCI algorithm. For all other parameters we use the default values as implemented in \texttt{tigramite}. Since we analyse warm and cool periods separately, each rolling two-year window spans non-contiguous calendar months. To handle this, we use LPCMCI's masking option with \texttt{mask\_type} = \texttt{'y'}, setting the mask to 1 for all months not belonging to the relevant warm or cool period. This ensures that causal dependencies are evaluated only at time steps within the seasonal period of interest, while masked observations can still serve as lagged causal ancestors.  We apply the same strategy to exclude missing load forecast values and the Texas freeze period (12–19 February 2021). In each case, we also mask the 7 days following the event to ensure that observations from these periods do not enter as lagged causes at up to $\tau_{\max}$. Although we decompose the electricity prices to their underlying components, system $\lambda$ and $\Delta_h$, we model all variables within a single causal graph to account for endogeneity between these components.

\paragraph{Incorporating domain knowledge} 
We incorporate domain knowledge into LPCMCI by specifying known relationships and ruling out implausible or forbidden connections based on economic theory and the temporal structure of the day-ahead market. This approach guides the algorithm by reducing the search space, thereby simplifying the detection and orientation of other causal relationships and mitigating the risk of spurious edges \citep{constantinou2023impact}. 

\textit{Renewable generation forecasts and weather} are treated as exogenous to market variables: they can cause prices, natural gas generation and load, but cannot be caused by them at any lag. This reflects the physical determination of weather and the fact that renewable forecasts are based on meteorological conditions rather than market outcomes. \textit{Forecast load} can cause market variables contemporaneously but not vice versa, reflecting the sequence of the day-ahead market in which load forecasts are formed before market clearing. At lagged timescales, market variables are permitted to cause forecast load, capturing potential demand responses to price signals. Contemporaneous relationships between forecast loads in different regions are restricted to bidirected edges, as regional loads are driven by common underlying factors (weather, economic activity) rather than causing each other within the same day. At longer lags, cross-regional load effects are unrestricted.

\textit{Natural gas hub prices and generation volumes} are subject to one additional restriction: they cannot cause day-ahead price components at lag zero, because day-ahead prices are determined on the afternoon of the preceding day while natural gas prices and generation volumes refer to the delivery day. At all other lags, gas variables follow the same assumptions as other market variables. Natural gas prices can be contemporaneously caused by weather but not vice versa.

\textit{Control variables} are constrained as follows: Seasonal harmonics can contemporaneously cause all endogenous variables but cannot be caused by them or share a latent common cause with them. The linear trend variable can cause market variables, load and renewable forecasts but has no relationship with weather. The weekend indicator can cause market variables and load but has no relationship with renewable forecasts or weather. These restrictions prevent the algorithm from fitting spurious causal pathways through deterministic controls. 

All relationships not explicitly constrained, including lagged effects between endogenous variables up to seven days, are left for the algorithm to determine from the data.

\paragraph{Interpretation of the output} 
As explained above, for each pair of variables $X^i_T$, $X^j_{T-s}$, $s \leq \tau$, LPCMCI performs the robust partial correlation test for every possible conditioning set, retaining the maximal p-value and the corresponding test statistic (i.e., the robust partial correlation coefficient). If this maximal $p$-value is smaller than a fixed hyperparameter $\alpha$, the association between the two variables is significant and cannot be explained by the remaining variables. Such significant  associations are encoded as edges in our causal graphs. %

Whether associations can be interpreted as causal effects is determined during the second phase of the algorithm. In particular, the causal graph we construct from the LPCMCI output contains four types of edges: directed edges ($\rightarrow$), bidirected edges ($\leftrightarrow$), partially directed edges ($\pardir$), and non-directed edges ($\nondir$). Due to the presence of latent variables, it is usually not possible to infer all causal directions, that is, to orient all edges, hence the latter two edge types. We interpret each edge type as follows, denoting by $A$ and $B$ any two random variables (nodes) in the graph, and not differentiating between direct and indirect causal effects \citep{spirtes2000causation, zhang2008causal, richardson2003causal}. 
\begin{itemize} \setlength{\itemsep}{0.3cm}
    \item[] ($A \rightarrow B$): $A$ has a causal effect on $B$.
    \item[] ($A \leftrightarrow B$): There is no causal effect of $A$ on $B$, and there is no causal effect of $B$ on $A$. Hence, there is a latent common cause of $A$ and $B$, which means that there is a latent variable $L$ such that $L$ has a causal effect on $A$ and $L$ has a causal effect on $B$.
    \item[] ($A \, \pardir \! B$): There is no causal effect of $B$ on $A$. This means that either $A$ has a causal effect on $B$ or there is a latent common cause of $A$ and $B$.
    \item[] ($A \nondir B$): There is a latent common cause of $A$ and $B$, or, either $A$ has a causal effect on $B$, or $B$ has a causal effect on $A$.
\end{itemize}

The absence of an edge suggests that any association between the two variables can be explained indirectly through other variables in the graph, providing evidence against a causal link. However, the lack of a link could also be due to other reasons. For example, the causal dynamics happen at a different temporal frequency and cannot be detected at the given frequency, or the causal effects are time-varying  with different strengths at different times and not stable across the full period. 

\clearpage
\section*{Generative AI and AI-assisted technologies in the writing process}\label{sec:declaration}
During the preparation of this work, the authors used ChatGPT and Claude to improve the readability and language of the manuscript and to assist with code development. After using these tools, the authors reviewed and edited the content as needed and take full responsibility for the content of the published article.

\clearpage
\section*{Supplementary material}
\setcounter{figure}{0}
\setcounter{table}{0}
\renewcommand{\thefigure}{S\arabic{figure}}
\renewcommand{\thetable}{S\arabic{table}}

\begin{figure}[htbp]
    \centering
    \includegraphics[width=0.8\textwidth]{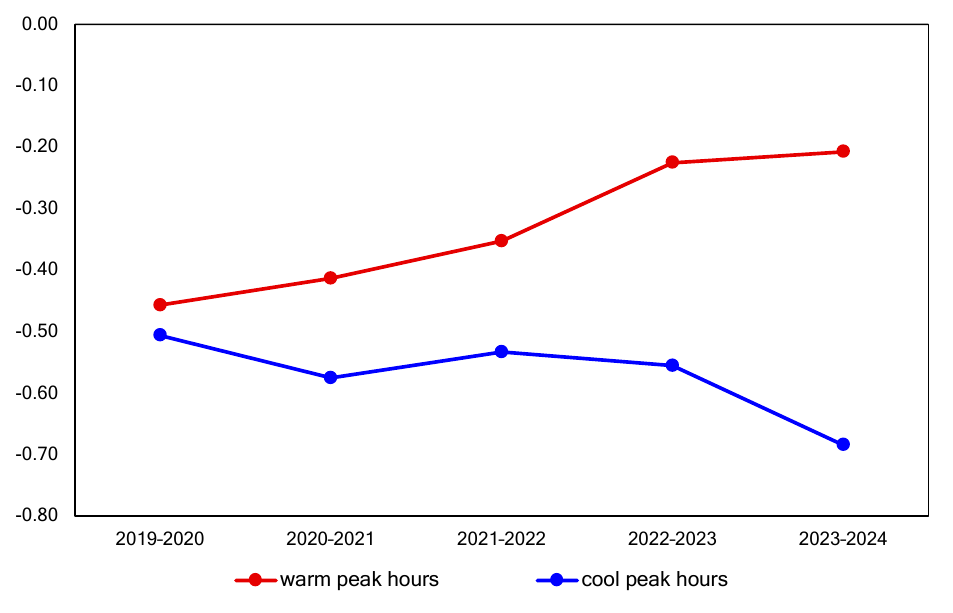}
    \caption{Partial correlation coefficient between Wind$_{west}$ and system $\lambda$ during peak hours.}
    \label{fig:figureS1}
\end{figure}

\begin{figure}[htbp]
    \centering
    \includegraphics[width=0.8\textwidth]{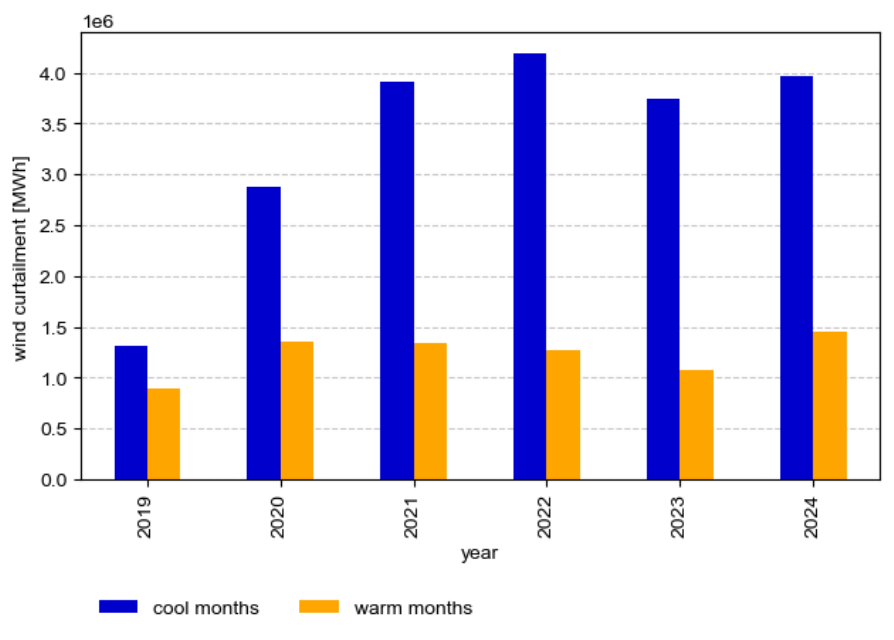}
    \caption{Wind curtailment volumes across warm and cool months, 2019–2024. Curtailment is calculated as the difference between high sustained limit (HSL) and base point (BP) from ERCOT's security-constrained economic dispatch (SCED) data, sourced via GridStatus.}
    \label{fig:figureS2}
\end{figure}

\clearpage
Here, we report the partial correlation coefficients of Wind$_{west}$ and natural gas prices on system lamda. For each period in which at least one gas hub causally affects system lambda, we computed the ratio of the absolute partial correlation of Wind$_{west}$ on lambda to the strongest natural gas hub effect on lambda. The 3.4 ratio reported in the main text is the mean across these periods. Periods with no detected causal effect from natural gas prices on lambda were excluded.\\

\begin{table}[ht]
\centering
\caption{Effect of Wind$_{west}$ on system $\lambda$ (partial correlation coefficient) by period.}
\label{tab:tableS1}
\renewcommand{\arraystretch}{1.5}
\begin{tabular}{| l | p{2cm} |p{2cm}| p{2cm}| p{2.5cm}|}
\hline
\textbf{Period} & \textbf{Peak Cool} & \textbf{Off-peak Cool} & \textbf{Peak Warm} & \textbf{Off-peak Warm} \\
\hline
\hline
        2019--2020 & $-$0.51 & $-$0.44 & $-$0.46 & $-$0.58\\
        \hline
        2020--2021 & $-$0.57 & $-$0.63 & $-$0.41 & $-$0.54\\
        \hline
        2021--2022 & $-$0.53 & $-$0.55  & $-$0.35 & $-$0.34\\
        \hline
        2022--2023 & $-$0.56 & $-$0.42 & $-$0.23 & $-$0.49\\
        \hline
        2023--2024 & $-$0.68 & $-$0.78  & $-$0.21 & $-$0.65\\

\hline
\end{tabular}
\end{table}

\begin{table}[ht]
\centering
\renewcommand{\arraystretch}{1.6}
\caption{Effect of natural gas hub prices on system $\lambda$ (partial correlation coefficient) by period.}
\label{tab:tableS2}
\begin{tabular}{| l | *{4}{p{2cm} r |}}
\hline
\textbf{Period} 
  & \multicolumn{2}{c|}{\textbf{Peak Cool}} 
  & \multicolumn{2}{c|}{\textbf{Offpeak Cool}} 
  & \multicolumn{2}{c|}{\textbf{Peak Warm}} 
  & \multicolumn{2}{c|}{\textbf{Offpeak Warm}} \\
\hline
\hline

2019--2020
  & \makecell[tl]{Waha (lag 1) \\ HH (lag 7)}         & \makecell[tr]{0.14 \\ 0.11}
  & \makecell[tl]{Waha (lag 1) \\ HH$_{Mc1}$ (lag 6)} & \makecell[tr]{0.15 \\ 0.14}
  & HSC (lag 1)                                        & 0.14
  & ---                                                & \\
\hline

2020--2021
  & Waha (lag 1) & 0.15
  & Waha (lag 1) & 0.14
  & ---          & 
  & Waha (lag 1) & 0.12 \\
\hline

2021--2022
  & ---                                                       & 
  & HH$_{Mc1}$ (lag 7)                                        & 0.12
  & HSC (lag 5)                                               & 0.18
  & \makecell[tl]{HH$_{Mc1}$ (lag 3) \\ HH$_{Mc1}$ (lag 7)}  & \makecell[tr]{0.15 \\ $-$0.12} \\
\hline

2022--2023
  & HH$_{Mc1}$ (lag 4) & 0.17
  & ---                & 
  & ---                & 
  & HH$_{Mc1}$ (lag 3) & 0.211 \\
\hline

2023--2024
  & ---          & 
  & HH (lag 2)   & 0.21
  & ---          & 
  & ---          & \\

\hline
\end{tabular}
\end{table}

\clearpage
\bibliography{literature}

\begin{thebibliography}{10}
\expandafter\ifx\csname url\endcsname\relax
  \def\url#1{\burl{#1}}\fi
\expandafter\ifx\csname urlprefix\endcsname\relax\def\urlprefix{URL }\fi
\providecommand{\bibinfo}[2]{#2}
\providecommand{\eprint}[2][]{\url{#2}}
\providecommand{\doi}[1]{\url{https://doi.org/#1}}
\bibcommenthead

\bibitem{bistline2021deep}
\bibinfo{author}{Bistline, J.~E.}, \bibinfo{author}{Roney, C.~W.},
  \bibinfo{author}{McCollum, D.~L.} \& \bibinfo{author}{Blanford, G.~J.}
\newblock \bibinfo{title}{Deep decarbonization impacts on electric load shapes
  and peak demand}.
\newblock \emph{\bibinfo{journal}{Environmental Research Letters}}
  \textbf{\bibinfo{volume}{16}}, \bibinfo{pages}{094054}
  (\bibinfo{year}{2021}).

\bibitem{shaffer2022changing}
\bibinfo{author}{Shaffer, B.}, \bibinfo{author}{Quintero, D.} \&
  \bibinfo{author}{Rhodes, J.}
\newblock \bibinfo{title}{Changing sensitivity to cold weather in texas power
  demand}.
\newblock \emph{\bibinfo{journal}{iScience}} \textbf{\bibinfo{volume}{25}}
  (\bibinfo{year}{2022}).

\bibitem{nerc_large_loads_2025}
\bibinfo{author}{{NERC}}.
\newblock \bibinfo{title}{Characteristics and risks of emerging large loads}.
\newblock \bibinfo{type}{Tech. Rep.}, \bibinfo{institution}{North American
  Electric Reliability Corporation} (\bibinfo{year}{2025}).
\newblock
  \urlprefix\url{https://www.nerc.com/globalassets/who-we-are/standing-committees/rstc/whitepaper-characteristics-and-risks-of-emerging-large-loads.pdf}.

\bibitem{norris_load_2025}
\bibinfo{author}{Norris, T.}, \bibinfo{author}{Profeta, T.},
  \bibinfo{author}{Patino-Echeverri, D.} \& \bibinfo{author}{Cowie-Haskell, A.}
\newblock \bibinfo{title}{Rethinking load growth: Assessing the potential for
  integration of large flexible loads in {US} power systems}.
\newblock \bibinfo{type}{Tech. Rep.}, \bibinfo{institution}{Duke University}
  (\bibinfo{year}{2025}).
\newblock \urlprefix\url{https://hdl.handle.net/10161/32077}.

\bibitem{heiss2026powering}
\bibinfo{author}{Heiss, A.}, \bibinfo{author}{Sautner, Z.} \&
  \bibinfo{author}{Schmid, T.}
\newblock \bibinfo{title}{{Powering AI: How Do Data Centers Affect Renewable
  Energy Investment?}}  (\bibinfo{year}{2026}).
\newblock \urlprefix\url{https://ssrn.com/abstract=6426079}.
\newblock \bibinfo{note}{Swiss Finance Institute Research Paper No. 26-29}.

\bibitem{peter_r_hartley_ercot_2024}
\bibinfo{author}{Hartley, P.~R.}, \bibinfo{author}{Medlock, K.~B., III} \&
  \bibinfo{author}{Hung, S.~Y.}
\newblock \bibinfo{title}{{ERCOT} and the {Future} of {Electric} {Reliability}
  in {Texas}}.
\newblock \bibinfo{type}{Tech. Rep.}, \bibinfo{institution}{Rice University’s
  Baker Institute for Public Policy} (\bibinfo{year}{2024}).
\newblock
  \urlprefix\url{https://www.bakerinstitute.org/sites/default/files/2024-01/BIPP-CES-ERCOT-Reliability-020724.pdf}.

\bibitem{tyler_hodge_data_2024}
\bibinfo{author}{{U.S. Energy Information Administration}}.
\newblock \bibinfo{title}{Data centers and cryptocurrency mining in {Texas}
  drive strong power demand growth}.
\newblock \bibinfo{howpublished}{U.S. Energy Information Administration,
  \textit{Today in Energy}} (\bibinfo{year}{2024}).
\newblock
  \urlprefix\url{https://www.eia.gov/todayinenergy/detail.php?id=63344}.
\newblock \bibinfo{note}{Accessed: 2025-11-14}.

\bibitem{eia_steo_2026}
\bibinfo{author}{{U.S. Energy Information Administration}}.
\newblock \bibinfo{title}{Short-term energy outlook data browser}
  (\bibinfo{year}{2026}).
\newblock
  \urlprefix\url{https://www.eia.gov/outlooks/steo/data/browser/#/?v=22}.
\newblock \bibinfo{note}{Accessed: 2026-02-25}.

\bibitem{woo2011impact}
\bibinfo{author}{Woo, C.-K.}, \bibinfo{author}{Horowitz, I.},
  \bibinfo{author}{Moore, J.} \& \bibinfo{author}{Pacheco, A.}
\newblock \bibinfo{title}{The impact of wind generation on the electricity
  spot-market price level and variance: The texas experience}.
\newblock \emph{\bibinfo{journal}{Energy Policy}}
  \textbf{\bibinfo{volume}{39}}, \bibinfo{pages}{3939--3944}
  (\bibinfo{year}{2011}).

\bibitem{woo2011wind}
\bibinfo{author}{Woo, C.-K.}, \bibinfo{author}{Zarnikau, J.},
  \bibinfo{author}{Moore, J.} \& \bibinfo{author}{Horowitz, I.}
\newblock \bibinfo{title}{Wind generation and zonal-market price divergence:
  Evidence from texas}.
\newblock \emph{\bibinfo{journal}{Energy Policy}}
  \textbf{\bibinfo{volume}{39}}, \bibinfo{pages}{3928--3938}
  (\bibinfo{year}{2011}).

\bibitem{ketterer2014impact}
\bibinfo{author}{Ketterer, J.~C.}
\newblock \bibinfo{title}{The impact of wind power generation on the
  electricity price in germany}.
\newblock \emph{\bibinfo{journal}{Energy economics}}
  \textbf{\bibinfo{volume}{44}}, \bibinfo{pages}{270--280}
  (\bibinfo{year}{2014}).

\bibitem{clo2015merit}
\bibinfo{author}{Cl{\`o}, S.}, \bibinfo{author}{Cataldi, A.} \&
  \bibinfo{author}{Zoppoli, P.}
\newblock \bibinfo{title}{The merit-order effect in the italian power market:
  The impact of solar and wind generation on national wholesale electricity
  prices}.
\newblock \emph{\bibinfo{journal}{Energy Policy}}
  \textbf{\bibinfo{volume}{77}}, \bibinfo{pages}{79--88}
  (\bibinfo{year}{2015}).

\bibitem{rintamaki2017does}
\bibinfo{author}{Rintam{\"a}ki, T.}, \bibinfo{author}{Siddiqui, A.~S.} \&
  \bibinfo{author}{Salo, A.}
\newblock \bibinfo{title}{Does renewable energy generation decrease the
  volatility of electricity prices? {A}n analysis of {D}enmark and {G}ermany}.
\newblock \emph{\bibinfo{journal}{Energy Economics}}
  \textbf{\bibinfo{volume}{62}}, \bibinfo{pages}{270--282}
  (\bibinfo{year}{2017}).

\bibitem{ajanaku2024comparing}
\bibinfo{author}{Ajanaku, B.~A.} \& \bibinfo{author}{Collins, A.~R.}
\newblock \bibinfo{title}{Comparing merit order effects of wind penetration
  across wholesale electricity markets}.
\newblock \emph{\bibinfo{journal}{Renewable Energy}}
  \textbf{\bibinfo{volume}{226}}, \bibinfo{pages}{120372}
  (\bibinfo{year}{2024}).

\bibitem{jkedrzejewski2022electricity}
\bibinfo{author}{J{\k{e}}drzejewski, A.}, \bibinfo{author}{Lago, J.},
  \bibinfo{author}{Marcjasz, G.} \& \bibinfo{author}{Weron, R.}
\newblock \bibinfo{title}{Electricity price forecasting: The dawn of machine
  learning}.
\newblock \emph{\bibinfo{journal}{IEEE Power and Energy Magazine}}
  \textbf{\bibinfo{volume}{20}}, \bibinfo{pages}{24--31}
  (\bibinfo{year}{2022}).

\bibitem{peters2016causal}
\bibinfo{author}{Peters, J.}, \bibinfo{author}{B{\"u}hlmann, P.} \&
  \bibinfo{author}{Meinshausen, N.}
\newblock \bibinfo{title}{Causal inference by using invariant prediction:
  identification and confidence intervals}.
\newblock \emph{\bibinfo{journal}{Journal of the Royal Statistical Society
  Series B: Statistical Methodology}} \textbf{\bibinfo{volume}{78}},
  \bibinfo{pages}{947--1012} (\bibinfo{year}{2016}).

\bibitem{cacciarelli2025we}
\bibinfo{author}{Cacciarelli, D.}, \bibinfo{author}{Pinson, P.},
  \bibinfo{author}{Panagiotopoulos, F.}, \bibinfo{author}{Dixon, D.} \&
  \bibinfo{author}{Blaxland, L.}
\newblock \bibinfo{title}{Do we actually understand the impact of renewables on
  electricity prices? {A} causal inference approach}.
\newblock \emph{\bibinfo{journal}{iEnergy}} \textbf{\bibinfo{volume}{4}},
  \bibinfo{pages}{247--258} (\bibinfo{year}{2025}).

\bibitem{tiedemann2024identifying}
\bibinfo{author}{Tiedemann, S.} \emph{et~al.}
\newblock \bibinfo{title}{Identifying elasticities in autocorrelated time
  series using causal graphs}.
\newblock \emph{\bibinfo{journal}{arXiv preprint arXiv:2409.15530}}
  (\bibinfo{year}{2024}).

\bibitem{mjelde2009market}
\bibinfo{author}{Mjelde, J.~W.} \& \bibinfo{author}{Bessler, D.~A.}
\newblock \bibinfo{title}{Market integration among electricity markets and
  their major fuel source markets}.
\newblock \emph{\bibinfo{journal}{Energy Economics}}
  \textbf{\bibinfo{volume}{31}}, \bibinfo{pages}{482--491}
  (\bibinfo{year}{2009}).

\bibitem{ferkingstad2011causal}
\bibinfo{author}{Ferkingstad, E.}, \bibinfo{author}{L{\o}land, A.} \&
  \bibinfo{author}{Wilhelmsen, M.}
\newblock \bibinfo{title}{Causal modeling and inference for electricity
  markets}.
\newblock \emph{\bibinfo{journal}{Energy Economics}}
  \textbf{\bibinfo{volume}{33}}, \bibinfo{pages}{404--412}
  (\bibinfo{year}{2011}).

\bibitem{nakajima2013testing}
\bibinfo{author}{Nakajima, T.} \& \bibinfo{author}{Hamori, S.}
\newblock \bibinfo{title}{Testing causal relationships between wholesale
  electricity prices and primary energy prices}.
\newblock \emph{\bibinfo{journal}{Energy Policy}}
  \textbf{\bibinfo{volume}{62}}, \bibinfo{pages}{869--877}
  (\bibinfo{year}{2013}).

\bibitem{park2006price}
\bibinfo{author}{Park, H.}, \bibinfo{author}{Mjelde, J.~W.} \&
  \bibinfo{author}{Bessler, D.~A.}
\newblock \bibinfo{title}{Price dynamics among {US} electricity spot markets}.
\newblock \emph{\bibinfo{journal}{Energy Economics}}
  \textbf{\bibinfo{volume}{28}}, \bibinfo{pages}{81--101}
  (\bibinfo{year}{2006}).

\bibitem{castagneto2014dynamic}
\bibinfo{author}{Castagneto-Gissey, G.}, \bibinfo{author}{Chavez, M.} \&
  \bibinfo{author}{Fallani, F. D.~V.}
\newblock \bibinfo{title}{Dynamic {G}ranger-causal networks of electricity spot
  prices: A novel approach to market integration}.
\newblock \emph{\bibinfo{journal}{Energy Economics}}
  \textbf{\bibinfo{volume}{44}}, \bibinfo{pages}{422--432}
  (\bibinfo{year}{2014}).

\bibitem{glymour2019review}
\bibinfo{author}{Glymour, C.}, \bibinfo{author}{Zhang, K.} \&
  \bibinfo{author}{Spirtes, P.}
\newblock \bibinfo{title}{Review of causal discovery methods based on graphical
  models}.
\newblock \emph{\bibinfo{journal}{Frontiers in Genetics}}
  \textbf{\bibinfo{volume}{10}} (\bibinfo{year}{2019}).

\bibitem{pearl2009causality}
\bibinfo{author}{Pearl, J.}
\newblock \emph{\bibinfo{title}{Causality}} \bibinfo{edition}{Second} edn
  (\bibinfo{publisher}{Cambridge University Press, Cambridge},
  \bibinfo{year}{2009}).
\newblock \bibinfo{note}{Models, reasoning, and inference}.

\bibitem{spirtes2000causation}
\bibinfo{author}{Spirtes, P.}, \bibinfo{author}{Glymour, C.} \&
  \bibinfo{author}{Scheines, R.}
\newblock \emph{\bibinfo{title}{Causation, prediction, and search}}
  \bibinfo{edition}{Second} edn.
\newblock Adaptive Computation and Machine Learning (\bibinfo{publisher}{MIT
  Press, Cambridge, MA}, \bibinfo{year}{2000}).

\bibitem{peters2017elements}
\bibinfo{author}{Peters, J.}, \bibinfo{author}{Janzing, D.} \&
  \bibinfo{author}{Sch\"{o}lkopf, B.}
\newblock \emph{\bibinfo{title}{Elements of Causal Inference: Foundations and
  Learning Algorithms}}, \bibinfo{pages}{205--208} (\bibinfo{publisher}{The MIT
  Press}, \bibinfo{year}{2017}).

\bibitem{shojaie2022granger}
\bibinfo{author}{Shojaie, A.} \& \bibinfo{author}{Fox, E.~B.}
\newblock \bibinfo{title}{Granger causality: A review and recent advances}.
\newblock \emph{\bibinfo{journal}{Annual Review of Statistics and Its
  Application}} \textbf{\bibinfo{volume}{9}}, \bibinfo{pages}{289--319}
  (\bibinfo{year}{2022}).

\bibitem{runge2023causal}
\bibinfo{author}{Runge, J.}, \bibinfo{author}{Gerhardus, A.},
  \bibinfo{author}{Varando, G.}, \bibinfo{author}{Eyring, V.} \&
  \bibinfo{author}{Camps-Valls, G.}
\newblock \bibinfo{title}{Causal inference for time series}.
\newblock \emph{\bibinfo{journal}{Nature Reviews Earth \& Environment}}
  \textbf{\bibinfo{volume}{4}}, \bibinfo{pages}{487--505}
  (\bibinfo{year}{2023}).

\bibitem{runge2019detecting}
\bibinfo{author}{Runge, J.}, \bibinfo{author}{Nowack, P.},
  \bibinfo{author}{Kretschmer, M.}, \bibinfo{author}{Flaxman, S.} \&
  \bibinfo{author}{Sejdinovic, D.}
\newblock \bibinfo{title}{Detecting and quantifying causal associations in
  large nonlinear time series datasets}.
\newblock \emph{\bibinfo{journal}{Science Advances}}
  \textbf{\bibinfo{volume}{5}}, \bibinfo{pages}{eaau4996}
  (\bibinfo{year}{2019}).

\bibitem{du2023renewable}
\bibinfo{author}{Du, P.}
\newblock \emph{\bibinfo{title}{Renewable Energy Integration for Bulk Power
  Systems: {ERCOT} and the Texas Interconnection}} Power Electronics and Power
  Systems (\bibinfo{publisher}{Springer International Publishing},
  \bibinfo{year}{2023}).
\newblock \urlprefix\url{https://link.springer.com/10.1007/978-3-031-28639-1}.

\bibitem{texas_counties_map}
\bibinfo{author}{{Texas Open Data Portal}}.
\newblock \bibinfo{title}{Texas counties map}.
\newblock
  \bibinfo{howpublished}{\url{https://data.texas.gov/dataset/Texas-Counties-Map/48ag-x9aa}}
  (\bibinfo{year}{2023}).
\newblock \bibinfo{note}{Accessed: 2025-11-04}.

\bibitem{eia860}
\bibinfo{author}{{U.S. Energy Information Administration}}.
\newblock \bibinfo{title}{Form {EIA}-860: {Annual Electric Generator Report},
  2019--2024}.
\newblock \urlprefix\url{https://www.eia.gov/electricity/data/eia860/}.

\bibitem{rrc_pdq}
\bibinfo{author}{{Railroad Commission of Texas}}.
\newblock \bibinfo{title}{Oil \& gas data query; general production query}.
\newblock
  \bibinfo{howpublished}{\url{https://webapps2.rrc.texas.gov/EWA/productionQueryAction.do}}
  (\bibinfo{year}{2025}).
\newblock \bibinfo{note}{Oil production by county, January--December 2024.
  Accessed: 2025-11-04}.

\bibitem{zarnikau2019market}
\bibinfo{author}{Zarnikau, J.}, \bibinfo{author}{Woo, C.~K.},
  \bibinfo{author}{Zhu, S.} \& \bibinfo{author}{Tsai, C.-H.}
\newblock \bibinfo{title}{Market price behavior of wholesale electricity
  products: Texas}.
\newblock \emph{\bibinfo{journal}{Energy Policy}}
  \textbf{\bibinfo{volume}{125}}, \bibinfo{pages}{418--428}
  (\bibinfo{year}{2019}).

\bibitem{madadkhani2024toward}
\bibinfo{author}{Madadkhani, S.} \& \bibinfo{author}{Ikonnikova, S.}
\newblock \bibinfo{title}{Toward high-resolution projection of electricity
  prices: A machine learning approach to quantifying the effects of high fuel
  and {CO2} prices}.
\newblock \emph{\bibinfo{journal}{Energy Economics}}
  \textbf{\bibinfo{volume}{129}}, \bibinfo{pages}{107241}
  (\bibinfo{year}{2024}).

\bibitem{gerhardus2020high-recall}
\bibinfo{author}{Gerhardus, A.} \& \bibinfo{author}{Runge, J.}
\newblock \bibinfo{title}{ in \textit{High-recall causal discovery for
  autocorrelated time series with latent confounders}} (eds
  \bibinfo{editor}{Larochelle, H.}, \bibinfo{editor}{Ranzato, M.},
  \bibinfo{editor}{Hadsell, R.}, \bibinfo{editor}{Balcan, M.} \&
  \bibinfo{editor}{Lin, H.}) \emph{\bibinfo{booktitle}{Advances in Neural
  Information Processing Systems}}, Vol.~\bibinfo{volume}{33}
  \bibinfo{pages}{12615--12625} (\bibinfo{publisher}{Curran Associates, Inc.},
  \bibinfo{year}{2020}).

\bibitem{weron_electricity_2014}
\bibinfo{author}{Weron, R.}
\newblock \bibinfo{title}{Electricity price forecasting: {A} review of the
  state-of-the-art with a look into the future}.
\newblock \emph{\bibinfo{journal}{International Journal of Forecasting}}
  \textbf{\bibinfo{volume}{30}}, \bibinfo{pages}{1030--1081}
  (\bibinfo{year}{2014}).

\bibitem{FellRoh2019}
\bibinfo{author}{Fell, H.} \& \bibinfo{author}{Roh, H.}
\newblock \bibinfo{title}{Wind energy, transmission, and production costs: Does
  increased connectivity help all?} (\bibinfo{year}{2019}).
\newblock
  \urlprefix\url{https://cenrep.ncsu.edu/cenrep/wp-content/uploads/2020/01/Roh_JMP.pdf}.
\newblock \bibinfo{note}{Job market paper, Accessed: 2026-01-19}.

\bibitem{eia_elec_browser}
\bibinfo{author}{{U.S. Energy Information Administration}}.
\newblock \bibinfo{title}{Electricity data browser: Net generation for all
  sectors, {T}exas, annual} (\bibinfo{year}{2024}).
\newblock \urlprefix\url{https://www.eia.gov/electricity/data/browser/}.

\bibitem{potomaceconomics2024}
\bibinfo{author}{{Potomac Economics}}.
\newblock \bibinfo{title}{2023 state of the market report for the {ERCOT}
  electricity markets}.
\newblock \bibinfo{type}{Tech. Rep.}, \bibinfo{institution}{Potomac Economics}
  (\bibinfo{year}{2024}).
\newblock
  \urlprefix\url{https://www.potomaceconomics.com/wp-content/uploads/2024/05/2023-State-of-the-Market-Report_Final_060624.pdf}.

\bibitem{eia_ng_2023}
\bibinfo{author}{{U.S. Energy Information Administration}}.
\newblock \bibinfo{title}{Difference in natural gas prices between {Texas} and
  {Henry Hub} narrowed in first half 2023}.
\newblock \bibinfo{howpublished}{U.S. Energy Information Administration,
  \textit{Today in Energy}} (\bibinfo{year}{2023}).
\newblock
  \urlprefix\url{https://www.eia.gov/todayinenergy/detail.php?id=60180}.

\bibitem{losada2020potential}
\bibinfo{author}{Losada~Carre{\~n}o, I.} \emph{et~al.}
\newblock \bibinfo{title}{Potential impacts of climate change on wind and solar
  electricity generation in {T}exas}.
\newblock \emph{\bibinfo{journal}{Climatic Change}}
  \textbf{\bibinfo{volume}{163}}, \bibinfo{pages}{745--766}
  (\bibinfo{year}{2020}).

\bibitem{navia_simon_power_2025}
\bibinfo{author}{Navia~Simon, D.} \& \bibinfo{author}{Diaz~Anadon, L.}
\newblock \bibinfo{title}{Power price stability and the insurance value of
  renewable technologies}.
\newblock \emph{\bibinfo{journal}{Nature Energy}}
  \textbf{\bibinfo{volume}{10}}, \bibinfo{pages}{329--341}
  (\bibinfo{year}{2025}).

\bibitem{prol2020cannibalization}
\bibinfo{author}{Prol, J.~L.}, \bibinfo{author}{Steininger, K.~W.} \&
  \bibinfo{author}{Zilberman, D.}
\newblock \bibinfo{title}{The cannibalization effect of wind and solar in the
  california wholesale electricity market}.
\newblock \emph{\bibinfo{journal}{Energy Economics}}
  \textbf{\bibinfo{volume}{85}}, \bibinfo{pages}{104552}
  (\bibinfo{year}{2020}).

\bibitem{eia2025ercot}
\bibinfo{author}{{U.S. Energy Information Administration}}.
\newblock \bibinfo{title}{{ERCOT} increasingly meets rising demand with solar,
  wind, and batteries}.
\newblock \bibinfo{howpublished}{U.S. Energy Information Administration,
  \textit{Today in Energy}} (\bibinfo{year}{2025}).
\newblock
  \urlprefix\url{https://www.eia.gov/todayinenergy/detail.php?id=66464}.

\bibitem{potomaceconomics2025}
\bibinfo{author}{{Potomac Economics}}.
\newblock \bibinfo{title}{2024 state of the market report for the {ERCOT}
  electricity markets}.
\newblock \bibinfo{type}{Tech. Rep.}, \bibinfo{institution}{Potomac Economics}
  (\bibinfo{year}{2025}).
\newblock
  \urlprefix\url{https://www.potomaceconomics.com/wp-content/uploads/2025/06/2024-State-of-the-Market-Report.pdf}.

\bibitem{ercot_da_ops_2024}
\bibinfo{author}{{Electric Reliability Council of Texas (ERCOT)}}.
\newblock \bibinfo{title}{Resources and {Day-Ahead} operations}.
\newblock \bibinfo{type}{Market Education Resource},
  \bibinfo{institution}{Electric Reliability Council of Texas}
  (\bibinfo{year}{2023}).
\newblock
  \urlprefix\url{https://www.ercot.com/files/docs/2023/07/25/2024_01%20Resource_DA_Ops.pdf}.

\bibitem{slusarewicz2018assessing}
\bibinfo{author}{Slusarewicz, J.~H.} \& \bibinfo{author}{Cohan, D.~S.}
\newblock \bibinfo{title}{Assessing solar and wind complementarity in {T}exas}.
\newblock \emph{\bibinfo{journal}{Renewables: Wind, Water, and Solar}}
  \textbf{\bibinfo{volume}{5}}, \bibinfo{pages}{1--13} (\bibinfo{year}{2018}).

\bibitem{gridstatusio}
\bibinfo{author}{{Grid Status}}.
\newblock \bibinfo{title}{{Gridstatusio Python Client}}.
\newblock \bibinfo{howpublished}{\url{https://www.gridstatus.io/}}.
\newblock \bibinfo{note}{Accessed: 2025-08--2025-10}.

\bibitem{ercot_data}
\bibinfo{author}{{Electric Reliability Council of Texas}}.
\newblock \bibinfo{title}{{Historical {DAM} Load Zone and Hub Prices
  ({NP4-180-ER})}}.
\newblock
  \bibinfo{howpublished}{\url{https://www.ercot.com/mp/data-products/data-product-details?id=NP4-180-ER}}.
\newblock \bibinfo{note}{Accessed: 2023-11 and 2025-01}.

\bibitem{eia_gridmonitor}
\bibinfo{author}{{U.S. Energy Information Administration}}.
\newblock \bibinfo{title}{{Hourly Electric Grid Monitor; Form EIA-930}}.
\newblock
  \bibinfo{howpublished}{\url{https://www.eia.gov/electricity/gridmonitor/knownissues/xls/ERCO.xlsx}}.
\newblock \bibinfo{note}{Accessed: 2025-08-09}.

\bibitem{Refinitiv}
\bibinfo{author}{{London Stock Exchange Group}}.
\newblock \bibinfo{title}{{LSEG (Refinitiv) Workspace}}.
\newblock \bibinfo{note}{Accessed: 2025-01 and 2025-10}.

\bibitem{menne2012overview}
\bibinfo{author}{Menne, M.~J.}, \bibinfo{author}{Durre, I.},
  \bibinfo{author}{Vose, R.~S.}, \bibinfo{author}{Gleason, B.~E.} \&
  \bibinfo{author}{Houston, T.~G.}
\newblock \bibinfo{title}{{An Overview of the Global Historical Climatology
  Network-Daily Database}}.
\newblock \emph{\bibinfo{journal}{Journal of Atmospheric and Oceanic
  Technology}} \textbf{\bibinfo{volume}{29}}, \bibinfo{pages}{897--910}
  (\bibinfo{year}{2012}).

\bibitem{ghcnd}
\bibinfo{author}{Menne, M.~J.} \emph{et~al.}
\newblock \bibinfo{title}{{Global Historical Climatology Network - Daily
  (GHCN-Daily), Version 3}}.
\newblock \bibinfo{note}{\url{https://doi.org/10.7289/V5D21VHZ}. Accessed:
  2025-09-02}.

\bibitem{sengupta2018national}
\bibinfo{author}{Sengupta, M.} \emph{et~al.}
\newblock \bibinfo{title}{{The National Solar Radiation Data Base (NSRDB)}}.
\newblock \emph{\bibinfo{journal}{Renewable and Sustainable Energy Reviews}}
  \textbf{\bibinfo{volume}{89}}, \bibinfo{pages}{51--60}
  (\bibinfo{year}{2018}).

\bibitem{granger1969investigating}
\bibinfo{author}{Granger, C. W.~J.}
\newblock \bibinfo{title}{Investigating causal relations by econometric models
  and cross-spectral methods}.
\newblock \emph{\bibinfo{journal}{Econometrica}} \textbf{\bibinfo{volume}{37}},
  \bibinfo{pages}{424--438} (\bibinfo{year}{1969}).

\bibitem{granger1980testing}
\bibinfo{author}{Granger, C.}
\newblock \bibinfo{title}{Testing for causality: A personal viewpoint}.
\newblock \emph{\bibinfo{journal}{Journal of Economic Dynamics and Control}}
  \textbf{\bibinfo{volume}{2}}, \bibinfo{pages}{329--352}
  (\bibinfo{year}{1980}).

\bibitem{mastakouri2021necessary}
\bibinfo{author}{Mastakouri, A.~A.}, \bibinfo{author}{Sch{\"o}lkopf, B.} \&
  \bibinfo{author}{Janzing, D.}
\newblock \bibinfo{editor}{Meila, M.} \& \bibinfo{editor}{Zhang, T.} (eds)
  \emph{\bibinfo{title}{Necessary and sufficient conditions for causal feature
  selection in time series with latent common causes}}.
\newblock (eds \bibinfo{editor}{Meila, M.} \& \bibinfo{editor}{Zhang, T.})
  \emph{\bibinfo{booktitle}{Proceedings of the 38th International Conference on
  Machine Learning}}, Vol. \bibinfo{volume}{139} of
  \emph{\bibinfo{series}{Proceedings of Machine Learning Research}},
  \bibinfo{pages}{7502--7511} (\bibinfo{publisher}{PMLR},
  \bibinfo{year}{2021}).

\bibitem{harris2013pc}
\bibinfo{author}{Harris, N.} \& \bibinfo{author}{Drton, M.}
\newblock \bibinfo{title}{P{C} algorithm for nonparanormal graphical models}.
\newblock \emph{\bibinfo{journal}{Journal of Machine Learning Research}}
  \textbf{\bibinfo{volume}{14}}, \bibinfo{pages}{3365--3383}
  (\bibinfo{year}{2013}).

\bibitem{liu2012highdimensional}
\bibinfo{author}{Liu, H.}, \bibinfo{author}{Han, F.}, \bibinfo{author}{Yuan,
  M.}, \bibinfo{author}{Lafferty, J.} \& \bibinfo{author}{Wasserman, L.}
\newblock \bibinfo{title}{High-dimensional semiparametric {G}aussian copula
  graphical models}.
\newblock \emph{\bibinfo{journal}{The Annals of Statistics}}
  \textbf{\bibinfo{volume}{40}}, \bibinfo{pages}{2293--2326}
  (\bibinfo{year}{2012}).

\bibitem{liu2009thenonparanormal}
\bibinfo{author}{Liu, H.}, \bibinfo{author}{Lafferty, J.} \&
  \bibinfo{author}{Wasserman, L.}
\newblock \bibinfo{title}{The nonparanormal: semiparametric estimation of high
  dimensional undirected graphs}.
\newblock \emph{\bibinfo{journal}{Journal of Machine Learning Research}}
  \textbf{\bibinfo{volume}{10}}, \bibinfo{pages}{2295--2328}
  (\bibinfo{year}{2009}).

\bibitem{constantinou2023impact}
\bibinfo{author}{Constantinou, A.~C.}, \bibinfo{author}{Guo, Z.} \&
  \bibinfo{author}{Kitson, N.~K.}
\newblock \bibinfo{title}{The impact of prior knowledge on causal structure
  learning}.
\newblock \emph{\bibinfo{journal}{Knowledge and Information Systems}}
  \textbf{\bibinfo{volume}{65}}, \bibinfo{pages}{3385--3434}
  (\bibinfo{year}{2023}).

\bibitem{zhang2008causal}
\bibinfo{author}{Zhang, J.}
\newblock \bibinfo{title}{Causal reasoning with ancestral graphs}.
\newblock \emph{\bibinfo{journal}{Journal of Machine Learning Research}}
  \textbf{\bibinfo{volume}{9}}, \bibinfo{pages}{1437--1474}
  (\bibinfo{year}{2008}).

\bibitem{richardson2003causal}
\bibinfo{author}{Richardson, T.} \& \bibinfo{author}{Spirtes, P.}
\newblock \bibinfo{title}{ in \textit{Causal inference via ancestral graph
  models}} (eds \bibinfo{editor}{Green, P.}, \bibinfo{editor}{Hjort, N.} \&
  \bibinfo{editor}{Richardson, S.}) \emph{\bibinfo{booktitle}{Highly Structured
  Stochastic Systems}} \bibinfo{pages}{83--105} (\bibinfo{publisher}{Oxford
  University Press}, \bibinfo{address}{USA}, \bibinfo{year}{2003}).

\end{thebibliography}

\end{document}